\documentclass[11pt]{article}
\usepackage[a4paper,hmarginratio=1:1,vmarginratio=2:3,totalwidth=15.6cm,
totalheight=21.1cm]{geometry}
\usepackage{bm,epstopdf,epsfig,amsmath,amssymb,amsfonts,
latexsym,comment,cancel,verbatim}
\usepackage[font=md,captionskip=8pt]{subfig}
\usepackage{amsmath}
\usepackage{amssymb}
\usepackage{amsfonts}
\usepackage{epsfig}

\renewcommand{\baselinestretch}{1.38}

\newcommand{\cO}{{\cal O}}

\newcommand{\hc}{\mbox{h.c.}}

\newcommand{\ra}{\rightarrow}
\newcommand{\be}{\begin{equation}}
\newcommand{\ee}{\end{equation}}
\newcommand{\bea}{\begin{eqnarray}}
\newcommand{\eea}{\end{eqnarray}}

\long\def\symbolfootnote[#1]#2{\begingroup
\def\thefootnote{\fnsymbol{footnote}}\footnote[#1]{#2}\endgroup} 
\begin{document}
 \begin{flushright}
OUTP-0930P\\
CERN-PH-TH-259
\end{flushright}

\thispagestyle{empty}

\vspace{1.2cm}

\begin{center}
{\Large {\bf Testing SUSY at the  LHC: \\
\medskip
Electroweak and Dark matter fine tuning at two-loop order.}}
\vspace{1.cm}

\textbf{S. Cassel$^{\,\,a}$, 
D.~M. Ghilencea$^{\,\,b,\!}$\symbolfootnote[1]{on leave from 
Theoretical Physics Department, IFIN-HH Bucharest MG-6, Romania.}, 
G. G. Ross$^{\,\,a,\,b,}$\symbolfootnote[2]{e-mail addresses:
s.cassel1@physics.ox.ac.uk, dumitru.ghilencea@cern.ch,
g.ross1@physics.ox.ac.uk}} \\[0pt]

\vspace{0.5cm} {$^a\, $ Rudolf\, Peierls\, Centre for Theoretical Physics,
\,University\, of\, Oxford,\\[0pt]
1 Keble Road, Oxford OX1 3NP, United Kingdom.}\\[6pt]
{\ $^b\, $ Department of Physics, CERN - Theory Division, CH-1211 Geneva 23,
Switzerland.}\\[6pt]
\end{center}

\def\baselinestretch{1.1}
\begin{abstract}
\noindent
In the framework of the Constrained Minimal Supersymmetric
Standard Model (CMSSM)  we evaluate the electroweak 
fine tuning measure
that provides a quantitative test of supersymmetry as a solution to 
the hierarchy problem. Taking account of current experimental constraints
we compute the fine
tuning at  two-loop order and determine the limits on the CMSSM
parameter space and the measurements at the LHC most relevant
in covering it. Without imposing the LEPII bound on the Higgs mass, it is shown
that the fine tuning computed at two-loop has a minimum
$\Delta=8.8$  corresponding to a Higgs mass $m_h=114\pm 2$~GeV.
Adding the constraint that the SUSY dark matter relic density should be within present bounds we find
$\Delta=15$ corresponding to $m_h=114.7\pm 2$~GeV and this rises to 
$\Delta=17.8$ ($m_h=115.9\pm 2$~GeV) for SUSY dark matter abundance  within 
3$\sigma$ of the
WMAP constraint.  
We extend the analysis to include the contribution of dark matter 
fine tuning. In this case  the overall
fine tuning and Higgs mass are only marginally larger for the case SUSY dark matter is subdominant and rises to $\Delta=28.7$ ($m_h=116.98\pm 2$~GeV) for the case of SUSY dark matter saturates the WMAP bound.
For a Higgs mass above these values,
fine tuning rises exponentially fast. The CMSSM spectrum that corresponds to minimal fine tuning
is computed and provides a benchmark for future searches. It is characterised by heavy squarks and sleptons
and light neutralinos, charginos and gluinos.
\end{abstract}

\newpage

\tableofcontents{}

\bigskip\bigskip
\section{Introduction}

With the Large Hadron Collider up and running, the search for
TeV-scale SUSY is now significantly closer.
Many models of  physics beyond the Standard Model (SM), in
particular its minimal  supersymmetric version (MSSM), will 
be directly tested. To do so, one has to quantify the viable range of
parameters entering in these models, which impact in particular 
 on the scale of its
low-energy supersymmetric spectrum. Previous theoretical and experimental
constraints such as those from LEP have already  tested a considerable 
amount of the MSSM parameter space and identified bounds on it.
To study  these bounds further 
it is instructive to investigate from a quantitative
perspective  the impact of the, so far, negative searches for  
low-energy supersymmetry. A quantitative measure of this impact is
the fine-tuning  measure $\Delta$ \cite{Ellis:1986yg,r3},
 that quantifies the degree of
cancellation between unrelated parameters that is needed to fix the
electroweak scale and can be extended to include the fine tuning
needed to obtain an acceptable dark matter abundance.
In this paper we shall perform such an investigation,
computing $\Delta$ at two-loop leading log
 order in the  Constrained MSSM (CMSSM).
For previous studies of the fine-tuning problem in a similar context
 see  \cite{Ellis:1986yg}-\cite{Cassel:2009cx}.

The electroweak fine-tuning measure,
 $\Delta$,  provides a measure of the probability 
of unnatural cancellations of 
soft masses in the expression of the 
electroweak scale $v^2\sim -\sum_i m_{soft,i}^2/\lambda$, 
($\lambda$ is Higgs quartic coupling) after including quantum corrections.
So $\Delta$ measures  the stability of 
the MSSM electroweak scale at the quantum level, 
with all available experimental and theoretical
constraints imposed. These include the LEP mass
bounds on supersymmetry masses, charge/colour breaking constraints, the
dark matter relic density constraint, and the measurement of $b\ra s \gamma$ and the anomalous
magnetic moment of the muon. In what follows we identify the constraints with the largest impact
on $\Delta$. We also extend the analysis to include the fine tuning, $\Delta^{\Omega}$, needed to satisfy the constraints on the SUSY dark matter abundance. The method can be
readily extended to other models that claim to solve the hierarchy
problem. For the case that the fine tuning is reduced by new states
with mass well above the TeV range, one may extend the analysis using
the effective Lagrangian in which the very heavy states have been
integrated out. For fine tuning and related issues in such
scenarios see for example
\cite{Cassel:2009ps,Dine,Antoniadis:2008es,Antoniadis:2009rn,Carena:2009gx},
where the MSSM Higgs mass can be increased nearer the LEP bounds 
by classical effects due to new physics beyond few TeV, which
ultimately reduces the fine tuning.

The  fine-tuning problem in the MSSM is  important 
not only for supersymmetry searches, but also for the Higgs physics since 
it is  intrinsically related to the value of the lightest Higgs mass
$m_h$, currently restricted by the LEPII  lower bound of  $114.4$ GeV
\cite{higgsboundLEP}.
As a result searches for $m_h$ are relevant to supersymmetry
phenomenology. In particular, the need to increase the SUSY prediction
for the Higgs mass by radiative corrections above the LEPII bound
means that the electroweak fine tuning measure rises exponentially with the Higgs mass.  If $\Delta$ becomes too large 
one can conclude that SUSY  fails to provide a solution to the
hierarchy problem. 
The parameter configuration (consistent with the non-observation of
SUSY states) that minimises $\Delta$ gives an indication of its most
likely values. We identify this configuration and investigate its
phenomenological implications. Also for a given upper  
value of $\Delta$ one can extract the corresponding range 
of parameter space of the CMSSM and of the superpartners masses.

In Section 2 we present the calculation of the electroweak fine tuning
 measure to two loop order. We are not aware of  a similar analysis of
 the fine tuning problem
at this level of precision (two-loop leading log), which is
responsible for a  $\Delta$ smaller than that usually found in the
literature.  In Section 3 we discuss the dependence of the fine tuning
on tan$\beta$ and the impact on the fine tuning coming from imposing
the bounds on the SUSY spectrum and the limits on $b\ra s \gamma$.
Using the dependence of the fine tuning measure on the parameters of
the CMSSM we then determine their allowed range consistent with a
given value of $\Delta$. This provides a quantitative measure of the
remaining parameter space range that remains to be tested. Next we
determine the dark matter abundance as a function of the fine tuning
measure showing that low fine tuning is consistent with acceptable
SUSY dark matter abundence. We conclude this Section by considering
the implications for the Higgs mass following from requiring low fine
tuning. We show that, without imposing the LEPII bound on the Higgs
mass,  $\Delta$ has a  minimum for a region of $m_h$ near the LEPII
bound.  In Section 4 we extend the fine tuning analysis to include the
fine tuning needed either to satisfy the dark matter bound or to
saturate the bound with SUSY dark matter. Finally in Section 5 we
discuss the predictions for the superpartner mass spectrum from the
fine tuning bound. We also determine the most likely spectrum that
minimises fine tuning and discuss
 the relative importance of  various LHC measurements
  in the test of the CMSSM.
Section 6 presents a summary and our conclusions.

\section{Computing
electroweak scale fine-tuning $\Delta$ at two-loop level}\label{section2}

In this section we present the strategy for evaluating
the MSSM fine-tuning at tree, one-loop
and two-loop (leading log) level; particular attention is paid to clarifying
the impact on fine tuning of the quantum corrections to couplings and masses.
With the standard two-Higgs doublet notation, the scalar potential
is \medskip
\begin{eqnarray} 
V&=&  m_1^2\,\,\vert H_1\vert^2
+  m_2^2\,\,\vert H_2\vert^2
- (m_3^2\,\,H_1 \cdot H_2+h.c.)\nonumber\\[6pt]
 &&
 ~+~
\frac{1}{2}\,\lambda_1 \,\vert H_1\vert^4
+\frac{1}{2}\,\lambda_2 \,\vert H_2\vert^4
+\lambda_3 \,\vert H_1\vert^2\,\vert H_2\vert^2\,
+\lambda_4\,\vert H_1\cdot H_2 \vert^2\nonumber\\[5pt]
 &&
 ~+~
\bigg[\,
\frac{1}{2}\,\lambda_5\,\,(H_1\cdot  H_2)^2+\lambda_6\,\,\vert H_1\vert^2\, 
(H_1 \cdot H_2)+
\lambda_7\,\,\vert H_2 \vert^2\,(H_1 \cdot H_2)+h.c.\bigg]
\label{2hdm}
\end{eqnarray}

\medskip\noindent
The couplings $\lambda_j$ and the soft masses receive
one- and two-loop corrections that for the MSSM are found
in \cite{Martin:1993zk,Carena:1995bx}. 
We shall use these results to evaluate the overall amount of 
fine-tuning of the electroweak
scale. Technical details of the procedure can  be found in the Appendix.

To evaluate the fine-tuning,  it is convenient to introduce the notation
\medskip
\begin{eqnarray}
m^2 &=&
 m_1^2 \, \cos^2 \beta +  m_2^2
 \, \sin^2 \beta - m_3^2 \, \sin 2\beta\nonumber\\[6pt]
\lambda &=&\frac{ \lambda_1^{} }{2} \, \cos^4 \beta 
+ \frac{ \lambda_2^{} }{2} \,  \sin^4 \beta 
+ \frac{ \lambda_{345}^{} }{4} \, \sin^2 2\beta 
+ \sin 2\beta \left( \lambda_6^{} \cos^2 \beta 
+ \lambda_7^{} \sin^2 \beta \right)\label{ml}
\end{eqnarray}

\medskip\noindent
with the assumption that, in the MSSM, at the UV scale
 $m_1^2=m_2^2=m_0^2+\mu_0^2$ while $m_3^2=B_0\,\mu_0$.
 The couplings  $\lambda_j^{}$ are assumed to be real 
and $\lambda_{345}^{} = \lambda_3^{} +\lambda_4^{} + \lambda_5^{}$.  
In\footnote{
When using the Yukawa couplings at the low energy scale as an input, 
expressed in terms of  the Higgs  vev and quark masses,
then $m$ and $\lambda$
pick up additional, implicit  dependence on $v=\langle h\rangle$ and
$\tan\beta$. Neglecting this dependence when evaluating 
$\Delta_p$ does not bring significant changes to the results 
for final $\Delta$.}
the MSSM, at  the tree level they are
\medskip
\bea
\lambda_1=\lambda_2=1/4 \,(g_1^2+g_2^2),\,\,\,\,\,
\lambda_3=1/4\, (g_2^2-g_1^2), \,\,\,\,\,
\lambda_4=-1/2\, g_2^2,\,\,\,\,\,
\lambda_{5,6,7}=0
\eea

\medskip\noindent
where $g_{1,2}$ are the $U(1)$ and $SU(2)$ gauge couplings
respectively.

The fine-tuning amount wrt to a set of parameters $\{p\}$ of the 
MSSM is then \cite{Ellis:1986yg,r3}
\medskip  
\begin{equation} 
\Delta \equiv \max \big\vert\Delta _{p}\big\vert_{p=\{\mu 
_{0}^{2},m_{0}^{2},m_{1/2}^{2},A_{0}^{2},B_{0}^{2}\}},
\qquad \Delta _{p}\equiv \frac{\partial \ln
  v^{2}}{\partial \ln p}  \label{ft} 
\end{equation}

\medskip\noindent
where all $p$ are input parameters at the UV scale of CMSSM, 
in the standard notation\footnote{One could also include 
in the set of parameters $p$, the top Yukawa coupling or the strong
coupling $\alpha_3$. For such parameters which are
measured, one can use the
modified fine tuning definition \cite{Ciafaloni:1996zh}
and with this, it turns out that in the cases we consider their
associated fine tuning never dominates.}
The minimisation of $V$ gives 
\bea
v^2=-m^2/\lambda,\qquad 
2 \lambda\frac{\partial m^2}{\partial \beta}=m^2\frac{\partial
  \lambda}{\partial \beta}
\eea
which fix $v$ and
$\beta$ as functions of the above MSSM bare parameters.
Taking into account that  $m^{2}=m^{2}(p,\beta )$,
 $\lambda =\lambda (p,\beta )$ we can find 
${\partial \beta }/{\partial p}$ from 
the second minimum condition for $V$. Using this, one finds\footnote{
Later the  min condition (fixing $\beta$) is used
to replace $B_0$ by $\tan\beta$ as an independent parameter.}
 \cite{Casas:2003jx}, see also \cite{Cassel:2009ps}: 
\medskip
\begin{eqnarray} \label{delta0}
\Delta _{p} &=& -\frac{p}{z}\,\bigg[\bigg(2
\frac{\partial ^{2}m^{2}}{\partial 
\beta ^{2}}+v^{2}\frac{\partial ^{2}\lambda }{\partial \beta ^{2}}\bigg)
\bigg(\frac{\partial \lambda }{\partial p}+\frac{1}{v^{2}}\frac{\partial 
m^{2}}{\partial p}\bigg)+\frac{\partial m^{2}}{\partial \beta }\frac{
\partial ^{2}\lambda }{\partial \beta \partial p}-\frac{\partial \lambda }{
\partial \beta }\frac{\partial ^{2}m^{2}}{\partial \beta \partial p}\bigg]
\nonumber
\\[6pt]
z &=& \lambda \,\bigg(2\frac{\partial ^{2}m^{2}}
{\partial \beta ^{2}}+v^{2}\frac{ 
\partial ^{2}\lambda }{\partial \beta ^{2}}\bigg)-\frac{v^{2}}{2}\,\bigg(
\frac{\partial \lambda }{\partial \beta }\bigg)^{2}
\end{eqnarray} 

\medskip\noindent
This result takes into account the dependence of $\beta$ on the MSSM
set of parameters ($p$).
This  formula  also takes into account the loop-effects to the 
quartic couplings as well as the $\tan\beta$ dependence of the
radiative corrections on the parameter ``p''. As we shall see later, 
such effects tend to 
reduce fine-tuning, in many cases rather 
significantly\footnote{The radiative corrections to couplings
$\lambda_j$ and soft masses $m_i$ bring about additional field
dependence, and therefore additional $v$ and $\beta$ dependence.
If we include the extra $v$
dependence, we find the fine-tuning $\Delta_p$ changes into 
$\Delta_p^{'}=\Delta_p/(1-\Delta_{v^2})$
with $\Delta_{v^2}$ {\it defined} by eq.(\ref{delta0}). 
 $\vert\Delta_{v^2}\vert\ll 1$ in most cases
examined. We do not include this effect here and
work with $\Delta$ defined by eqs.(\ref{ft}), (\ref{delta0}).}.

For comparison with similar studies 
of fine-tuning, a comment is in place here.
After some algebra, one can show that
the general formula of $\Delta$ (eqs.(\ref{ft}), (\ref{delta0})) reduces, 
in the limit of removing the 
loop corrections to quartic couplings $\lambda_i$, $i=1,2...$, to the  
more familiar ``master formula''  \cite{Dimopoulos:1995mi}
(see also \cite{Chankowski:1997zh,Chankowski:1998xv})
\medskip
\bea
\Delta_{p}&=&\frac{p}{(\tan^2\beta-1)\,m_Z^2}\,
\bigg\{
\frac{\partial  m_1^2}{\partial p}-\tan^2\beta\,\frac{\partial
  m_2}{\partial p}\nonumber\\[7pt]
&&\qquad\qquad
-\,\,\frac{\tan\beta}{\cos
  2\beta}\bigg[1+\frac{m_Z^2}{m_1^2+\tilde m_2^2}\bigg]
\bigg[2\frac{\partial m_3^2}{\partial p}-\sin
  2\beta\,\Big(\frac{\partial m_1^2}{\partial p}+\frac{\partial
     m_2^2}{\partial p}\Big)
\bigg]\bigg\}\label{gd}
\eea

\bigskip\noindent
This formula is sometimes used
as  the starting point in  works that evaluate
electroweak scale fine-tuning, by using in it the loop corrected soft masses. 
However, for accurate estimates, it is necessary to
take full account of radiative corrections, as done by eqs.(\ref{ft}),
(\ref{delta0}). Indeed, the loop corrections to the
quartic couplings significantly reduce the amount of fine-tuning, in
some cases by a factor as  large as  2, and these corrections are not
accounted for by eq.(\ref{gd}). This can be seen by considering the
one-loop correction $\delta$ to $\lambda_2\ra \lambda_2(1+\delta)$, due to
stop/top Yukawa couplings. Usually $\delta=\cO(1)$. Including it one finds
(for details see eq.(26) in \cite{Cassel:2009ps}):
\medskip
\bea
\Delta_{p}\propto \frac{p}{(1+\delta)\,m_Z^2}
+\cO(1/\tan\beta), \qquad p=\mu_0^2,
m_0^2, m_{1/2}^2, A_0^2, B_0^2.
\eea

\bigskip\noindent
This is showing that one-loop corrections to the quartic 
coupling reduce the amount of fine-tuning significantly. In fact it is
the smallness of the quartic Higgs coupling (fixed in the MSSM by
gauge interactions) that is at the origin of substantial 
tree-level fine tuning.
That this is so can be seen from the relation $v^2\sim -m_i^2/\lambda$
where $v$ is of $\cO(100)$ GeV, $m_i\sim \cO(TeV)$ while at the same time
$\lambda<1$, which makes it difficult to separate the EW and SUSY
breaking scales (for a discussion see \cite{Giudice:2006sn}).
Loop corrections increase the quartic  couplings and in most cases
reduce the overall amount of fine tuning.

For a general two-Higgs doublet model formulae (\ref{ft}),
(\ref{delta0}) can be expressed in terms of only derivatives of
couplings and of masses wrt to the corresponding parameter,
see Appendix in \cite{Cassel:2009ps} (also Appendix A.1). 
For the CMSSM we
use this result, in which we consider the full two-loop (leading log)
corrections to the
quartic couplings and masses. This defines unambiguously our procedure
for evaluating the EW fine-tuning in CMSSM.

\section{Electroweak fine tuning and its effects on the
Higgs mass}\label{section3}

In the following, the numerical results we present for 
$\Delta$ include two-loop corrections with:

\medskip
\noindent
$\bullet$  radiative  electroweak breaking (EWSB),

\noindent
$\bullet$  non-tachyonic SUSY 
particle masses (avoiding colour and charge breaking (CCB) vacua). 

\noindent
$\bullet$ experimental constraints considered: bounds on superpartner
masses, electroweak precision data, $b\ra s\,\gamma$, $b\ra \mu\,\mu$
and anomalous magnetic moment $\delta a_\mu$, as detailed in
 Table~\ref{contable}.

\noindent
$\bullet$ consistency of $m_h$ with the LEPII bound (114.4 GeV) and/or 
consistency with thermal relic density constraint, only 
if stated explicitly.

\bigskip
\begin{table}[htdp]
\begin{center} 
\begin{tabular}{|c|c|} \hline
Constraint & Reference \\ \hline
SUSY particle masses & Routine in MicrOmegas 2.2,  
``MSSM/masslim.c" \\
$\delta a_\mu^{} < 366 \times 10^{-11}$ &  PDG 
(sys. and stat. $1\sigma$ errors added linearly) \\
$3.20  < 10^{4} ~ \mbox{Br}(b \to s \gamma) < 3.84$ & PDG 
(sys. and stat. $1\sigma$ errors added linearly)
\\
$\mbox{Br} (b \to \mu \mu) < 1.8 \times 10^{-8}$ &  Particle Data Group \\
$-0.0007< \delta \rho < 0.0012$ &  Particle Data Group \\ \hline
\end{tabular}
\end{center}
\caption{\small Constraints tested using MicrOMEGAs 2.2 with
 SuSpect 2.41 spectrum calculator.
 Particle Data Group: http://pdg.lbl.gov/.}
\label{contable}
\end{table}

\medskip\noindent
Using these constraints  we evaluated $\Delta$ numerically.
\begin{figure}[t]
\center
\def\baselinestretch{1.1}
\includegraphics[width=9.cm]{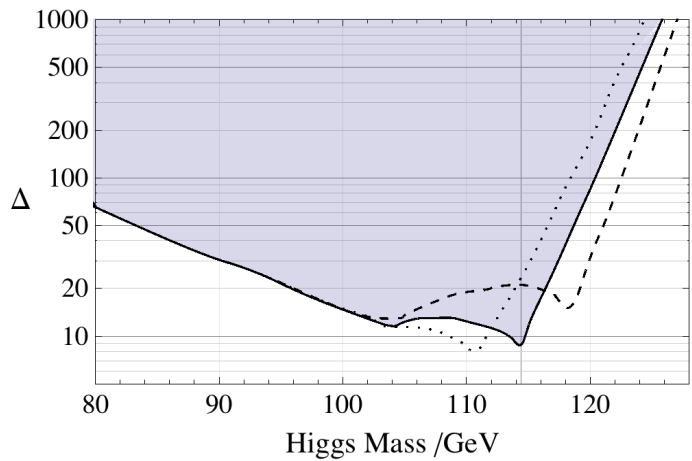}
\caption{\small
Fine tuning vs Higgs mass, in a two-loop analysis.
 The data points are for $2 \leq \tan \beta \leq 55$.
The solid line is the minimum fine tuning with central values 
$(\alpha_3, m_t)=(0.1176,173.1$\,GeV).  The dashed line corresponds to
 $(\alpha_3, m_t)=(0.1156,174.4$\,GeV) and the dotted line to
$(0.1196,171.8$\,GeV), to account for  1$\sigma$ experimental errors in
$\alpha_3$ and top mass \cite{TeV}.
This is the ``worst'' case scenario, when such deviations combine such as
to give the largest variation of $\Delta$.
An increase of $\alpha_3(m_Z)$ or reduction of
 $m_t(m_Z)$ by 1$\sigma$ have similar effects, which can be also
 understood from the relation between the mass of top evaluated 
 at $m_Z$ and at $m_t$. 
Keeping either $\alpha_3$ or $m_t$ fixed to its central value
and varying the other within
1$\sigma$ brings a curve situated half-way between the continuous line
and the corresponding dashed or dotted line.
 The LEPII bound of $114.4$~GeV is indicated by a vertical line. 
Note the steep ($\approx$ exponential) increase of $\Delta$ 
on both sides of 
its minimum value situated near the LEPII bound.}\label{2loop}
\end{figure}
The  LEPII bound on the mass of the Higgs provides  an important
constraint for the MSSM  since it requires quantum corrections in order
to be satisfied. Large quantum corrections need in turn large soft
masses, which in turn trigger large fine-tuning. This is seen
from the loop corrections to $m_h$ which give a strong exponential
dependence on $m_h$,  $\Delta \sim m_{0}^2\sim \exp(m_h^2/m_{top}^2)$.
To examine this dependence in detail, in the following we 
choose to present the numerical results as a function of $m_h$.
 Unless stated otherwise, the LEPII bound on $m_h$ is not imposed.
The relic density constraint is  imposed only after all the
constraints other than  the LEPII bound on $m_h$ are
satisfied and, when done, this is stated explicitly.

Before proceeding to present our numerical results obtained with the
above constraints, let us mention the details of the procedure
followed. First the fine-tuning is evaluated at two-loop order, including the
dominant third generation supersymmetric threshold effects to the
scalar potential. The scan is done over all parameter space using 
a slightly simplified two-loop calculation performed by a Mathematica
code, based on the formulae in the Appendix, and optimised to run
quickly. For the points in phase space that have the smallest
fine-tuning (say with $\Delta<1000$), the analysis is re-done 
using (the slower) SOFTSUSY 3.0.10 \cite{Allanach:2001kg} 
that includes all the two-loop radiative effects mentioned.
This two-step procedure is extremely important, since otherwise the CPU
run time using SOFTSUSY alone would be about 
6 years (when run on 30 parallel processors at 3GHz each),
which prevented previous investigations at this precision level.
Our two-loop analysis is also important because there is a
significant difference between one-loop and two-loop values for
overall $\Delta$, and  was not performed in the past.
QCD effects can compete at two-loop   with Yukawa couplings effects,
can dominate them and also displace the minimum of 
the fine tuning from its one-loop value.
 Regarding the Higgs mass, its value 
computed with SOFTSUSY agrees with that found using
SuSpect \cite{Djouadi:2002ze}  within $0.1$ GeV, but can 
differ by $\pm 2$ GeV \cite{Allanach:2004rh} 
from the value found using FeynHiggs \cite{Hahn:2009zz}.
We use the SOFTSUSY Higgs mass for  all figures in the paper.
Given the  small discrepancy with FeynHiggs, coming from
higher order terms in the perturbative expansion, the LEPII bound
should be interpreted as $m_h>114.4\pm 2$ GeV. In the following 
analysis the results are always quoted with respect
 to the central value of $m_h$.
\begin{figure}[t]
\center
\def\baselinestretch{1.1}
\includegraphics[width=10cm]{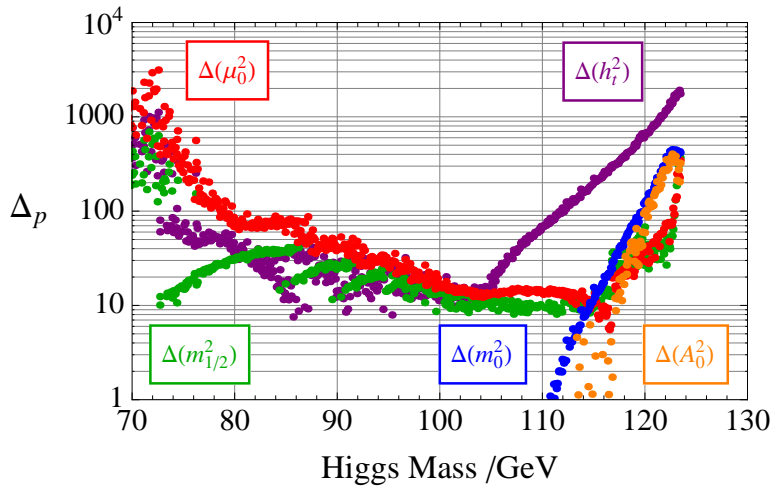}
\caption{\small
The plot displays the various contributions $\max\vert\Delta_p\vert$,
$p=\mu_0^2,m_0^2, A_0^2, B_0^2, m_{1/2}^2$, to the electroweak fine-tuning
$\Delta$ presented in Figure~\ref{2loop}. The largest of these for all
$m_h$ gives the curve presented in Figure~\ref{2loop}.
At low $m_h$, $\Delta_{\mu_0^2}$ (red) is dominant, while at large
$m_h$, $\Delta_{m_0^2}$ (blue) is dominant (with $\Delta_{A_0^2}$
reaching similar values near 120 GeV).
The transition between the two regions is happening at about 114.5
GeV. Note that in this plot the LEPII bound is not imposed at any time.
Although $\Delta_{h_t^2}$ (purple) is presented above for illustration, 
this contribution is always sub-dominant when
 assuming the  modified definition of fine-tuning \cite{Ciafaloni:1996zh},
appropriate for  measured parameters (as we do in the text).}\label{dplot}
\end{figure}

Turning now to  the numerical results, Figure~\ref{2loop} 
presents the two-loop result for the dependence of overall 
electroweak fine tuning $\Delta$ as a function of the Higgs mass. 
The dark matter constraint and the LEPII bound on $m_h$ are not included.
The loop effects reduce the fine tuning amount; the dominant
effects come from quantum corrections to the quartic couplings (rather
than to soft masses), which
are increased by radiative effects and thus reduce $\Delta$.
$\Delta$ is seen to have a  minimum close to the LEPII
bound of $m_h$. The individual contributions to $\Delta$ are shown in
Figure~\ref{dplot}. Below the LEPII bound, detailed calculations show that 
the minimal value of $\Delta$ is dominated by $\Delta_{\mu_0^2}$ 
and this increases rapidly for decreasing $m_h$.
For values of $m_h$ above the LEPII bound, $\Delta$ is  dominated 
by\footnote{Larger $m_h$ requires larger $m_{\tilde t}^2\sim m_0^2$,
and a larger $m_0$, above the focus point region,  increases
$\Delta_{m_0^2}$.} $\Delta_{m_0^2}$.
This happens at the edge of the focus point region.
 The transition from the dominant
$\Delta_{\mu_0^2}$ regime to the dominant $\Delta_{m_0^2}$ regime
occurs near the LEPII bound value, and this is the point where 
the QCD radiative effects  become important.
 This can be seen from Figure~\ref{2loop}
 where an increase by 1$\sigma$ of  $\alpha_3$ corresponds to a larger
 $\Delta$ for a same, fixed value of $m_h$.
 In this sense one could even say that
the minimal value of $\Delta$ is situated
at the transition region between dominant effects, 
Yukawa versus QCD interactions.
Away from the minimum of $\Delta$, fine 
tuning increases dramatically, roughly exponentially. 
This is because, as discussed above, $\Delta$ depends exponentially
on~$m_h$.

In conclusion, the fine tuning at two-loop  with all the latest 
constraints is minimised for
\medskip
\bea
&&\Delta\approx 8.8,\qquad m_h=114\pm 2\,\,\, \textrm{GeV} 
\eea

\medskip\noindent
with the theoretical uncertainty of $\pm 2$ GeV,  explained earlier.

Note that our analysis also investigated the contribution to 
$\Delta$ coming from the uncertainty in measured ``parameters'' 
such as top Yukawa and strong coupling. Using the modified\footnote{$\bar{\Delta}_p= \Delta_p \times (\sigma_p / p)$ where $\sigma_p$ is the 1$\sigma$ error in the parameter $p$ derived from experimental observation.} definition
\cite{Ciafaloni:1996zh}, appropriate for measured parameters, 
we find their fine-tuning is sub-dominant.

Finally, let us also mention
that the  reduction of fine-tuning that we have
seen is mostly  due to (two-) loop corrections considered,
 particularly to quartic
couplings and is actually very significant, given  the conservative
scenario considered here assuming
universal gaugino mass structure of the CMSSM; relaxing this 
condition could reduce \cite{Kane:1998im}  $\Delta$ further.

\subsection{Constraints on $\Delta$ from fixing $\tan\beta$}

\begin{figure}[t]
\center
\def\baselinestretch{1.1}
\includegraphics[width=9.9cm] {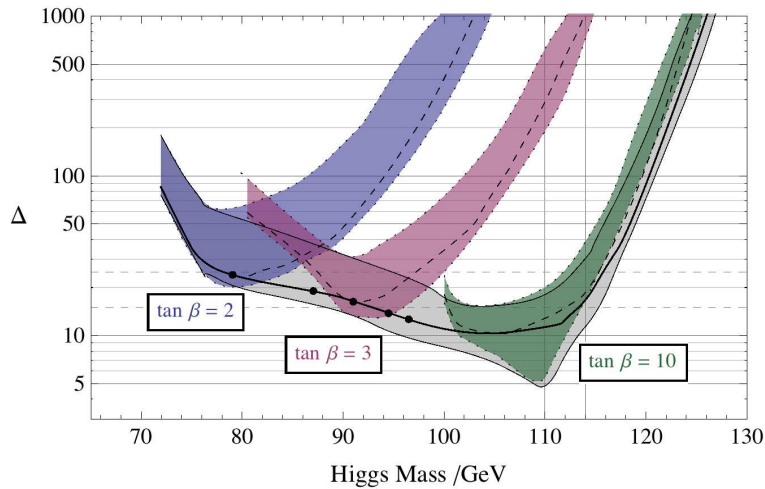}
\caption{\small
Minimum fine tuning versus Higgs mass at two-loop, showing the influence of
loop effects and $\tan \beta$ on fine tuning. All constraints listed
in Table~\ref{contable} are included. The upper and lower lines
associated with the coloured regions are the 1-loop without
thresholds for $\lambda$ and soft masses and ``full" 1-loop
results respectively (similarly for the grey region, for all
$\tan\beta$). The minimum 2-loop fine tuning  is found between these
two cases. The solid lines refer to the scan $2 \leq \tan \beta \leq
55$. The black points give the positions of minimum $\Delta$ 
for fixed $\tan \beta$ from 2 to 4 inclusive in steps of 0.5.}
\label{loopP}
\end{figure}

It is interesting to examine the fine-tuning
for fixed values of some of the parameters present, 
to see the individual impact of such constraints on $\Delta$. Here we do 
this for a fixed value of $\tan\beta$.
This is shown\footnote{
In Figures~\ref{loopP}, \ref{chargcon}, \ref{bsgcon} only, 
$\Delta$ is computed with our Mathematica code instead of SOFTSUSY,
due to long CPU time constraints. This explains the small difference in
shape between the two-loop line in these three figures from the more 
accurate one in Figure~\ref{2loop} and all other figures (based on SOFTSUSY).}
 in Figure~\ref{loopP} for
 increasing vales of
$\tan\beta$, with $\tan\beta=2$ (blue), $\tan\beta=3$ (red), $\tan\beta=
10$ (green).  Increasing $\tan\beta$ 
shifts the curves of $\Delta$ towards larger $m_h$ and, to a limited extent,
to lower fine-tuning, reached for medium $\tan\beta \sim 10$, also for
larger $\tan\beta\sim 40$ (see later, Fig.\ref{all} (a)).
As may be seen from the 
figure, the two-loop expressions for
soft masses and couplings bring values of $\Delta$ which 
are situated between the higher ``tree-level'' curve (i.e. 
tree-level for $\lambda_i$, one-loop for soft masses without field
dependent threshold effects)  and the lower one-loop
case (one-loop for both $\lambda_i$ and soft masses).
This is consistent with what one would expect from a convergent,
perturbative, loop expansion. The roughly exponential 
behaviour of $\Delta$ at large $m_h$ for fixed $\tan\beta$
can bring a significant variation of $\Delta$.

\subsection{Constraints on $\Delta$ from the SUSY spectrum}

\begin{figure}[t!]
\center
\def\baselinestretch{1.1}
\includegraphics[width=9.9cm] {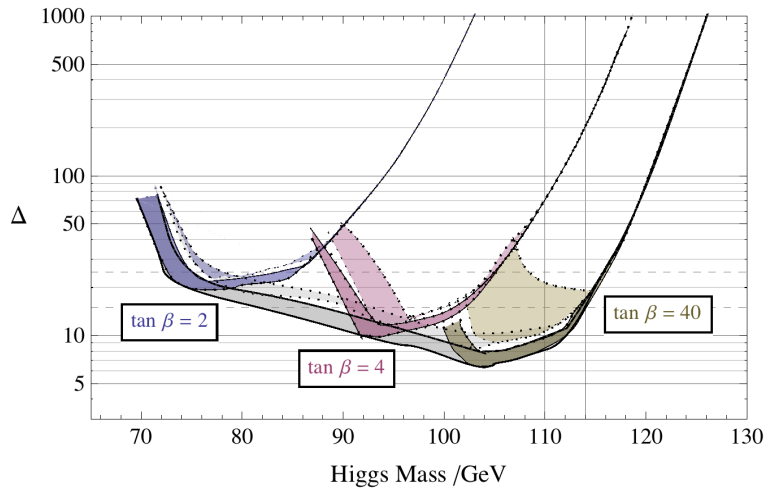}
\caption{\small
Minimum fine tuning vs Higgs mass, showing the influence of mass
constraints on fine tuning. The results are at 2-loop with the upper
shaded (coloured) areas connecting the case of only applying 
the SUSY spectrum constraints (lower line) to that with all 
constraints listed in Table~\ref{contable} (upper line).
The lower shaded (coloured) areas connect the
cases of only applying a chargino lower mass limit of 80 and 94 GeV
for the lower and upper lines respectively.
The results for the scan $2 \leq \tan \beta \leq 55$ are also shown
by the grey shaded area, 
with similar convention for upper/lower continuous lines delimiting it.
}\label{chargcon}
\end{figure}

Here we examine the impact on $\Delta$ due to
constraints related to the supersymmetric spectrum.
The key features of the impact of this spectrum on $\Delta$ can be
seen from the limits on the chargino mass considered  in
Figure~\ref{chargcon}. Currently, the (lightest) chargino mass bound
is the most important, followed by that of the neutralino. It also
turns out that the gluino mass limit is not very constraining.
These results follow recent experimental data, since
using the 1998 data it was the neutralino mass bound that was more
constraining for~$\Delta$.

In Figure~\ref{chargcon}, the effect of the chargino mass
$m_{\chi_1}>94$ and $m_{\chi_1}>80$ GeV is shown by the upper
and lower continuous curves. While these can have some impact on
fine-tuning for values of $m_h$ already ruled out experimentally,
for $m_h>114.4$ GeV, the effect is overlapping that from $b\ra s \gamma$
(see later).
Further, the close vicinity of the upper and lower dotted curves
corresponding to all constraints in Table~\ref{contable} and
to the SUSY spectrum limits respectively shows that the latter are
the main constraints at the moment for $\Delta$ at low $\tan \beta$. 
The graph presents $\Delta$ and $m_h$ computed  using quartic 
couplings and mass expressions evaluated at two-loop.

\subsection{Constraints on $\Delta$ from $b\ra s\gamma$}

\begin{figure}[t!]
\center
\def\baselinestretch{1.1}
\includegraphics[width=9.9cm]{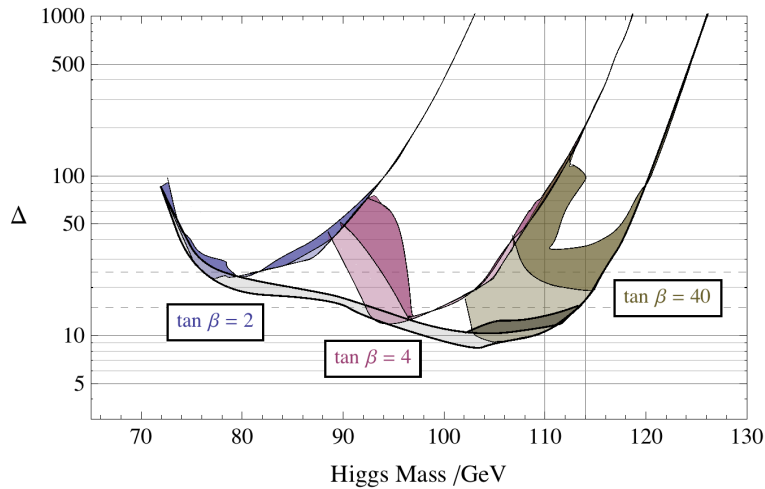}
\caption{\small
Minimum fine tuning vs Higgs mass, showing the influence of the $b \to
s \gamma$ constraint.  The results are at 2-loop with the lighter
shading connecting the case of only applying the SUSY spectrum
constraints (lower line) to that with also the $b \to s \gamma$
constraint listed in Table~\ref{contable} (upper line). This upper
edge of this shading is indistinguishable from the solid line which
includes all constraints in Table \ref{contable}. The darker shading
extends up to the minimum fine tuning limits for the case, $3.52 <
10^{4}\, \mbox{Br}(b \to s \gamma) < 3.77$, with the other constraints
as given in Table~\ref{contable}. The results for the scan $2 \leq
\tan \beta \leq 55$ are also shown, between the two continuous 
and almost parallel lower curves.}\label{bsgcon}
\end{figure}

\begin{figure}[t!]
\center
\def\baselinestretch{1.1}
\includegraphics[width=9.9cm]{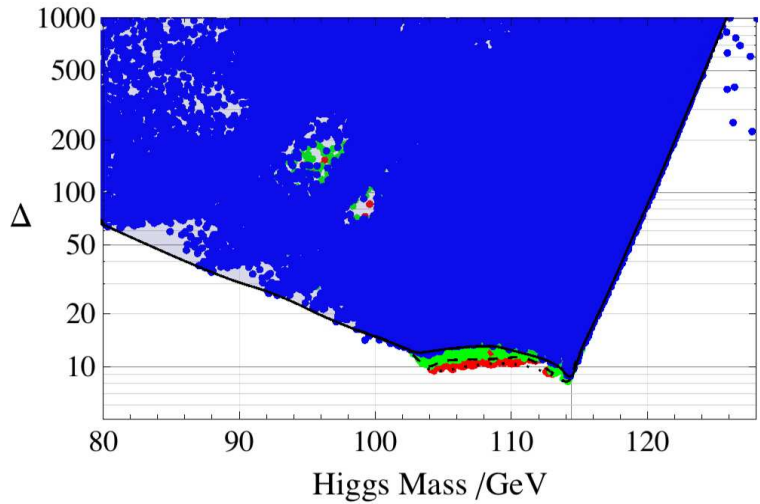}
\caption{\small
Minimum fine tuning vs Higgs mass, showing the influence of the $b \to
s \gamma$ constraint.
In the presence of this constraint with 3$\sigma$ limits, the red (lower) region is removed. Within the 1$\sigma$ limits, the green (middle) band is removed leaving the blue (upper) points.
This is a two-loop leading 
log approximation, obtained using SOFTSUSY.
}\label{bsgamma}
\end{figure}

Figure~\ref{bsgcon} gives the impact of the $b\ra s\gamma$ constraint 
on $\Delta$. The lower limit of the $ b \to s \gamma$ constraint for a given
coloured area (fixed $\tan\beta$) restricts the
right hand edges of the plot, while the upper limit restricts
its left hand side. These curves also depend on the mass limits - these
 constraints are not fully independent.
For the experimentally allowed area of $m_h>114.4$ GeV, the impact of
the constraint $b\ra s\gamma$ is rather small; in this case its
effect is overlapping that of the SUSY mass limits, as can be seen from
the rhs of the plots for individual $\tan\beta$ plots. The combination
of the SUSY mass limits and $b\ra s\gamma$ constraint currently
dominate the restriction on how small the fine tuning could be, see also
Figure~\ref{bsgamma}. In this last figure one can easily see, at
two-loop,  the impact on $\Delta$ of removing the $b\ra s \gamma$ constraint.
For a related analysis of $b\ra s\gamma$ see recent \cite{Nath}.

The other constraints listed in Table~\ref{contable} do eliminate further
mSUGRA points, but have a negligible 
effect on the fine tuning limits.
With the current mass limits, a change in the $\delta a_\mu^{}$ constraint by
factors of 2 or more does not affect these results significantly.

\subsection{Constraints on $\Delta$ and the CMSSM parameters}\label{cmssm}

\begin{figure}[th!]
\center
\def\baselinestretch{1.1}
\subfloat[Fine Tuning vs $\tan \beta$]{\includegraphics[width=6.6cm]{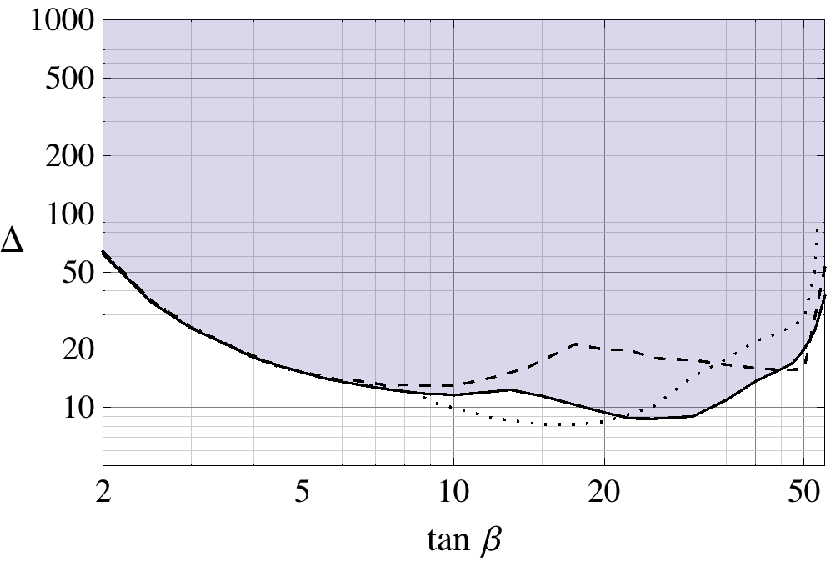}}
\hspace{4mm}
\subfloat[Fine Tuning vs $\tan \beta$]{\includegraphics[width=6.6cm]{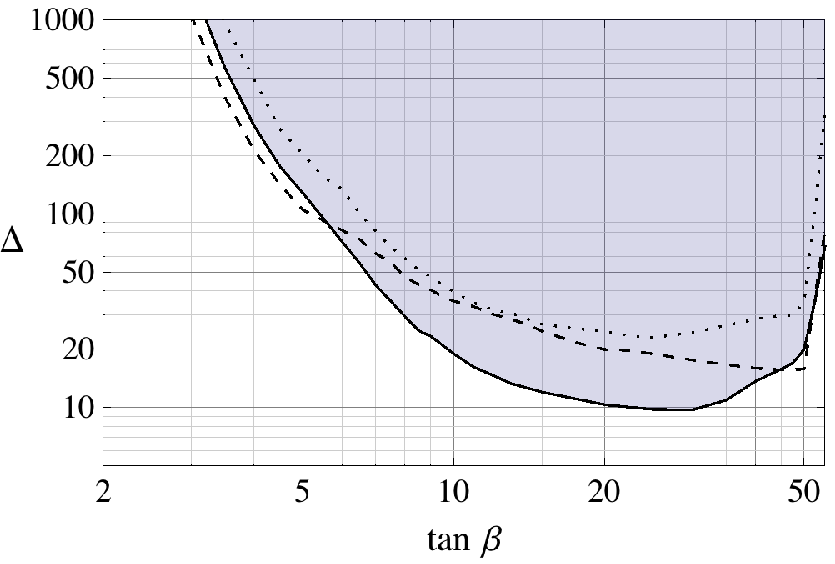}}
\hspace{4mm}
\subfloat[Fine Tuning vs
$A_0^{}$]{\includegraphics[width=6.6cm]{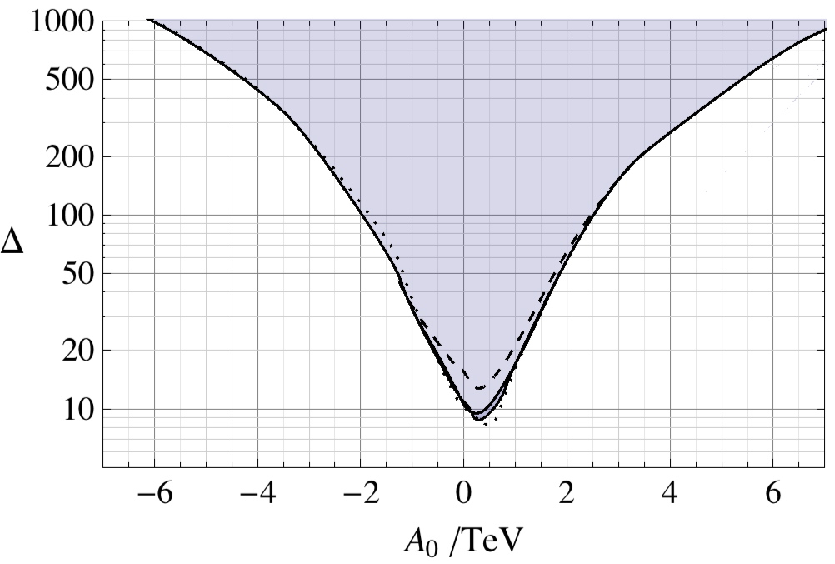}}
\hspace{4mm}
\subfloat[Fine Tuning vs $m_0^{}$]{\includegraphics[width=6.6cm]{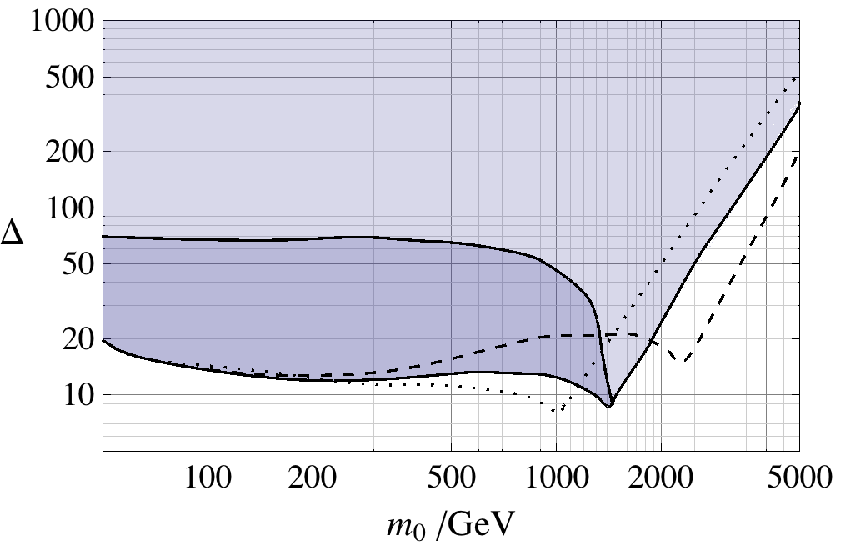}} 
\hspace{4mm}
\subfloat[Fine Tuning vs
$m_{1/2}^{}$]{\includegraphics[width=6.6cm]{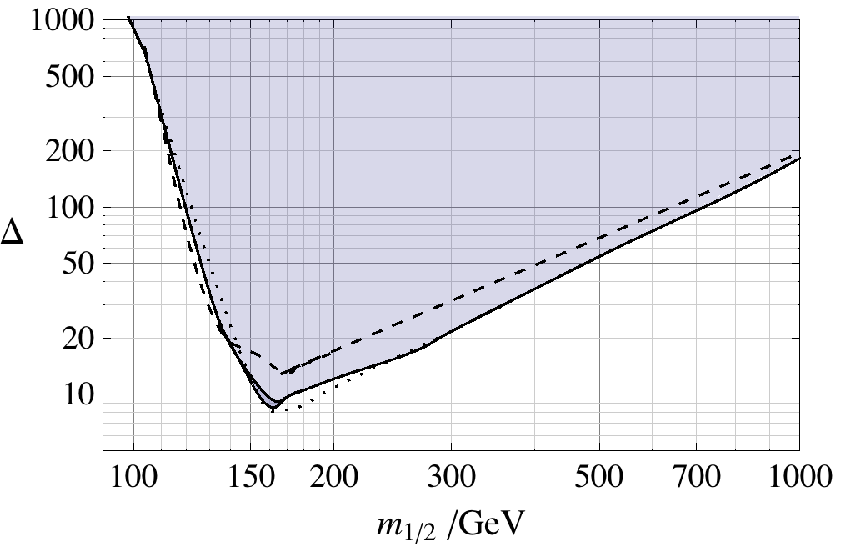}}
\hspace{4mm}
\subfloat[Fine Tuning vs $\mu$]{\includegraphics[width=6.6cm]{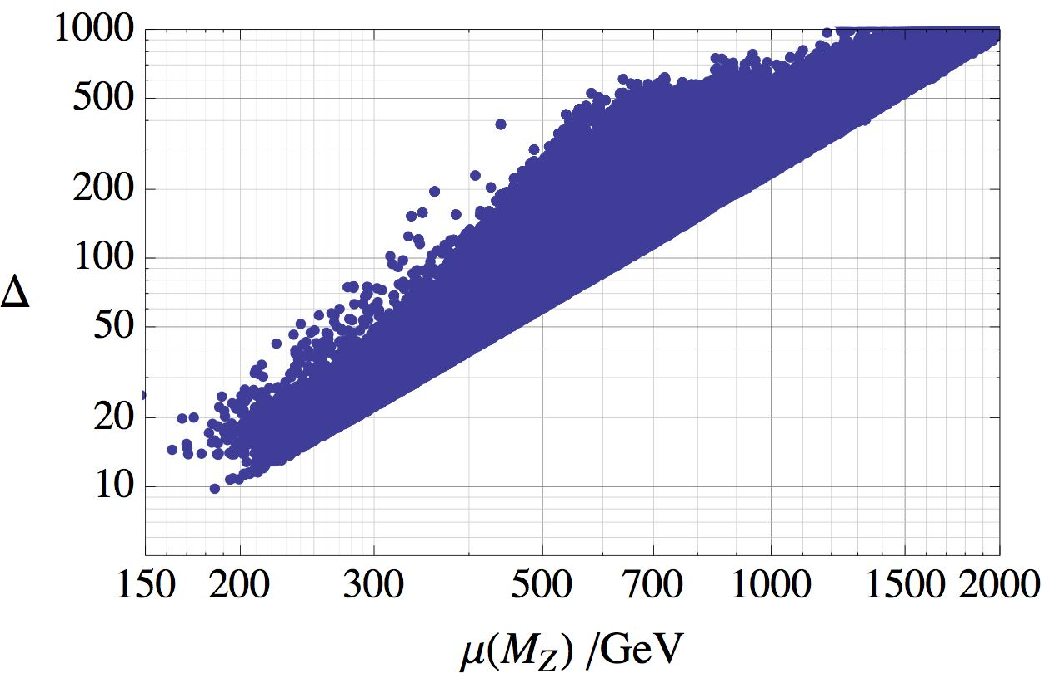}}
\caption{{\protect\small Dependence of minimum fine tuning on SUSY
parameters ($\protect\mu >0$, relic density unrestricted). The solid, 
dashed and dotted lines are as explained in Fig~\protect\ref{2loop}. 
No bound on $m_h$ is applied in figure (a). 
In (c), (d), (e), the darker shaded regions are eliminated when
 $m_{h}^{{}}>114.4$~GeV is applied for the case with the central
 $\left( \protect\alpha _{3}^{{}},m_{t}^{{}}\right) $ values. 
In (b) and (f), $m_h > 114.4$\,GeV is applied, and the points in (f)
 are only for the central $\left( \protect\alpha
 _{3}^{{}},m_{t}^{{}}\right) $ values.}}\label{all}
\end{figure}

The fine tuning measure can be easily applied to establish the
remaining allowed range for the MSSM SUSY parameters. In
Figure~\ref{all} we plotted the dependence of the total fine tuning 
wrt various parameters. To understand some aspects of the
dependence of the electroweak scale fine tuning on the MSSM
parameters,  let us use, for the sake of discussion,
one of the two minimum conditions  which, when ignoring quantum 
corrections to quartic couplings, simplifies to:
\begin{eqnarray}\label{mu0fixtree}
\frac{m_Z^2}{2} &=&  \frac{{m}_{1}^2-{m}_{2}^2 
\tan^2 \beta}{\tan^2 \beta -1}
\end{eqnarray}
%
\noindent
Using  2-loop RGE solutions at $\tan \beta =10$
(for details see the Appendix), one has
\smallskip
\begin{eqnarray}
{m}_1^2 \,(m_Z^{}) &\approx& 0.99\, \mu_0^2 + 0.946\, m_0^2 
+ 0.331\, m_{1/2}^2 + 0.044\, A_0^{} \, m_{1/2}^{} - 0.013\, A_0^2
\label{m1uv}
\\
{m}_2^2 \,(m_Z^{}) &\approx& 0.99\, \mu_0^2 - 0.080\, m_0^2 
- 2.865\, m_{1/2}^2 + 0.445\, A_0^{} \, m_{1/2}^{} - 0.099\, A_0^2
\label{m2uv}
\end{eqnarray} 

\smallskip\noindent
It is the large cancellation between the $\mu_0^2$ and $m_{1/2}^2$ terms
that is often  responsible for the large fine tuning (note however
that  this argument ignores the impact of quantum corrections to
quartic couplings, known to reduce the fine-tuning).
This leads to  the approximate relation
 $\Delta_{\mu_0^2}^{} \sim  \Delta_{m_{1/2}^2}^{}$. As
low fine tuning prefers small $\mu_0^{}$, small $m_{1/2}^{}$ is also
preferred and this is observed in Fig~\ref{all} (e), (f). The rise in
fine tuning at small $m_{1/2}$ is a consequence of the constraints
such as the chargino mass limit.

The near flat distribution of minimum fine tuning in $m_0^{}$ is a
result of the coefficient of $m_0^{}$ in $m_2^{}$ being driven close
to zero. The fine tuning with respect to $m_0^{}$ then rarely
dominates, until we reach values of $m_h$ above the LEPII bound ($m_0$
at the edge of focus point region). The result of applying the Higgs
mass constraint also excludes a region with small $m_{1/2}$ at
$m_0^{}$ below 1.5~TeV.
 The focus point at $m_0^{} \sim 1.5$~TeV where the minimum
 of $m_{1/2}$ is possible, corresponds to the point 
where fine tuning is minimised. This only occurs for large $\tan \beta$,
and this available ``dip" in fine tuning in the mSUGRA space
disappears  as $\tan \beta$ is reduced.

Figure~\ref{all}(c) indicates that a small trilinear coupling 
$|A_0^{}| \lesssim 1$~TeV is preferred for the smallest
fine tuning. This follows from a similar argument for preferring
small $m_{1/2}$. Increasing $|A_0^{}|$ requires larger cancellations
with $\mu$ to set the electroweak scale. However, once the Higgs mass
constraint is applied, $A_0^{}$ is driven negative for small $\tan \beta$ in order to
maximise the stop mixing. The related increase in the minimum fine
tuning from being in this region of parameter space then
follows. This is important for small $\tan \beta$ where the
tree level Higgs mass is smallest.
 The sign structure of the UV
 parameter coefficients in $m_2^{}$ leads to a preference in a
 small, positive $A_0^{}$.

As mentioned earlier, the fine tuning measure $\Delta$
can be used to  establish the remaining parameter space of the CMSSM
compatible with a solution to the hierarchy problem. Assuming that
$\Delta=100$ is the upper limit beyond which we consider that SUSY
failed to solve the hierarchy problem, we obtain the following bounds:
\smallskip
\begin{equation}\label{cmssmparameters}
\begin{tabular}{rlcrcl}
$m_{h}$ & $<~121$~\mbox{GeV} & \hspace{5mm} & $5.5~<$ & $\tan \beta $
  & 
$<~55
$ \\
$\mu $ & $<~680$~\mbox{GeV} &  & 120~\mbox{GeV}$~<$ & $m_{1/2}$ & $<~720$~
\mbox{GeV} \\
$m_{0}$ & $<~3.2$~\mbox{TeV} &  & $-2.0$~\mbox{TeV}$~<$ & $A_{0}$ & $<~2.5$~
\mbox{TeV}
\end{tabular}
\end{equation}

\smallskip\noindent
These values can be easily re-calculated for a different value of
$\Delta$.

\subsection{Constraints on $\Delta$ from the relic density}\label{cmssmrelic}

\begin{figure}[!th]
\center
\def\baselinestretch{1.1}
\begin{tabular}{cc|cr|} 
\subfloat[Fine tuning vs $A_0$]{\includegraphics[width=6.35cm]{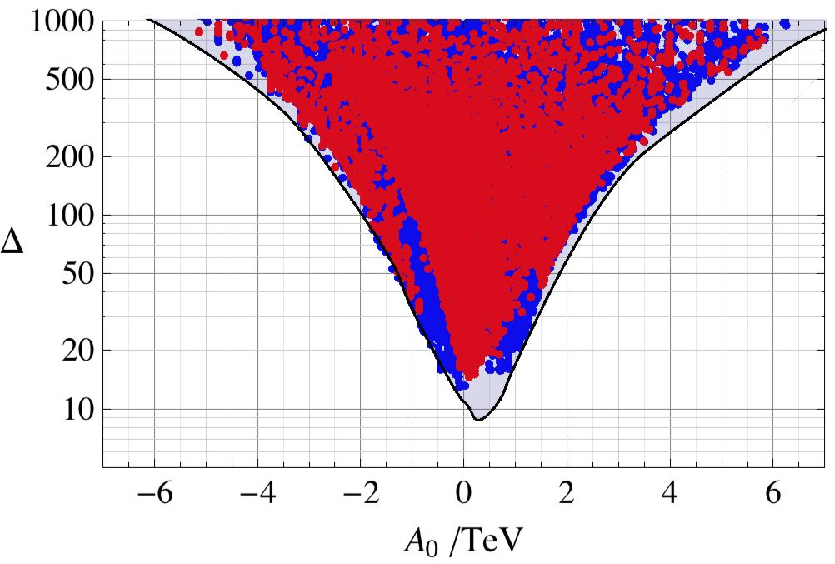}}
\hspace{3mm}
\subfloat[\!Fine tuning vs $A_0$, $m_h\!>\!114.4$GeV]{
\includegraphics[width=6.4cm]{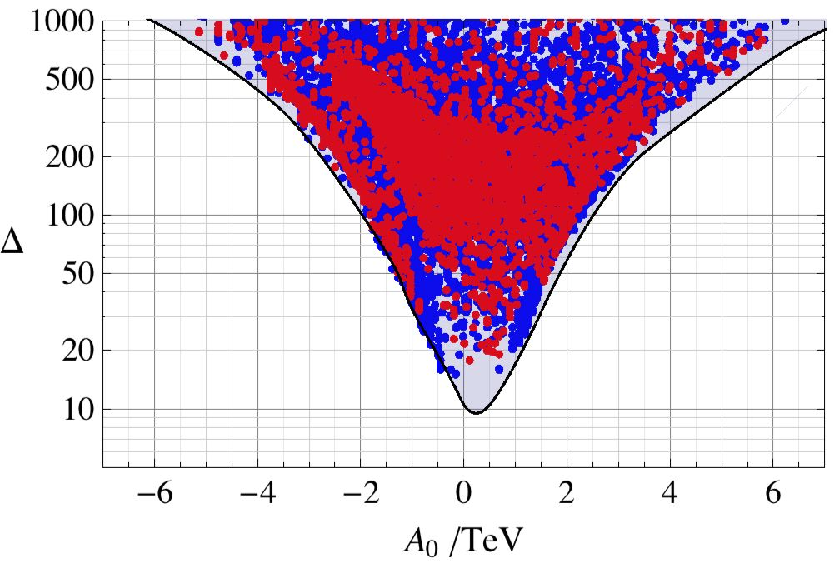}}
\end{tabular}
\begin{tabular}{cc|cr|} 
\subfloat[Fine tuning vs $m_0^{}$]{\includegraphics[width=6.5cm]{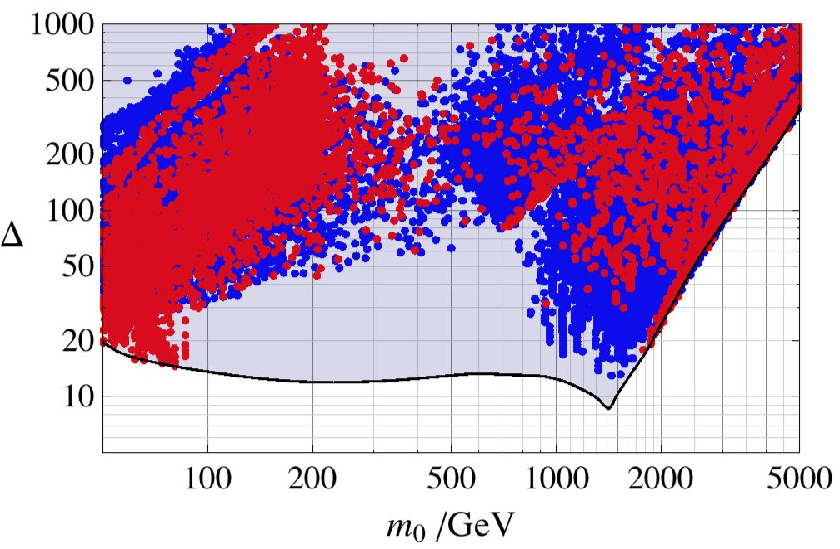}} 
\hspace{3mm}
\subfloat[\!Fine tuning vs $m_0^{}$, $m_h\!>\!114.4$\,GeV]{
\includegraphics[width=6.5cm]{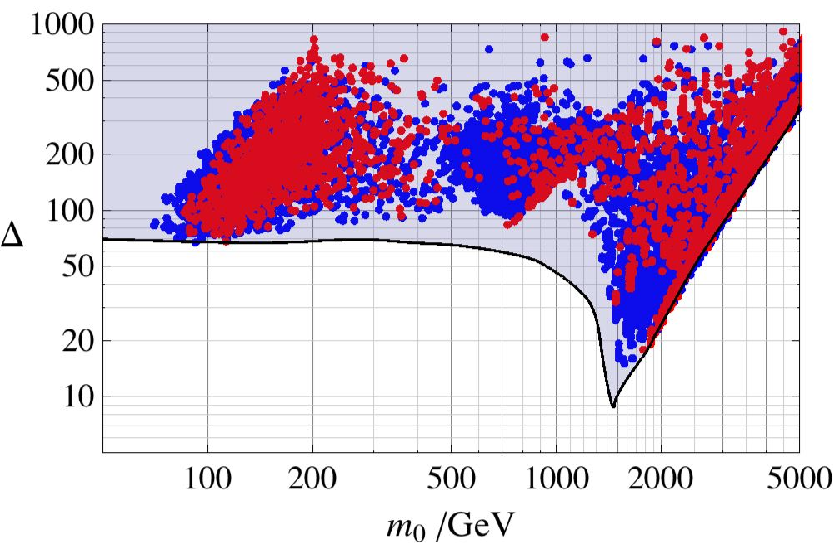}}
\end{tabular}
\begin{tabular}{cc|cr|} 
\subfloat[Fine 
tuning vs $m_{1/2}$]{\includegraphics[width=6.5cm]{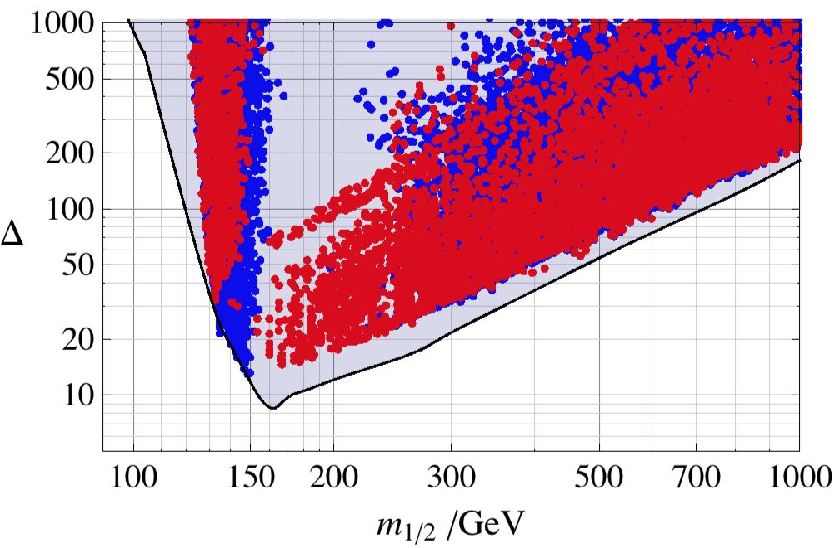}}
\hspace{3mm}
\subfloat[\!Fine tuning vs $m_{1/2}$, $m_h\!>\!114.4$\,GeV]{
\includegraphics[width=6.5cm]{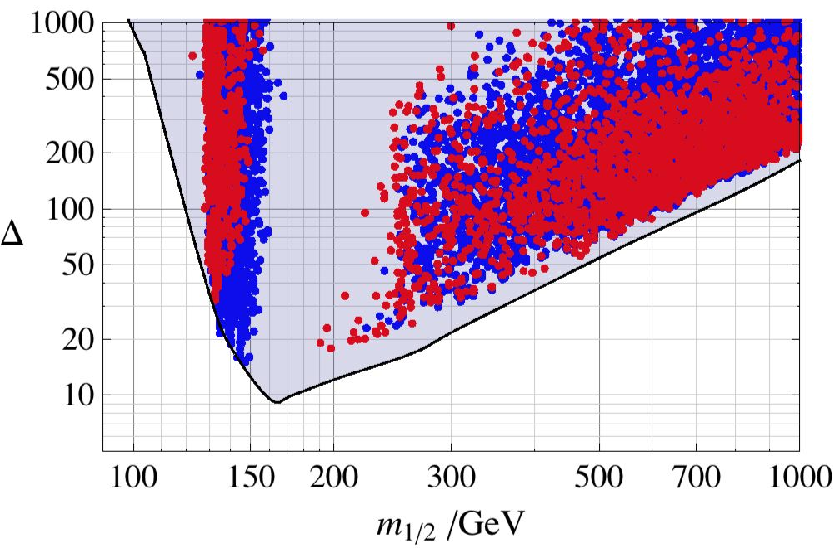}}
\end{tabular}
\caption{\protect\small 
Dependence of minimum fine tuning on SUSY
parameters, with the relic density saturated
 within 3$\sigma$ of the WMAP bound (in red). 
The WMAP bound is $\Omega h^2=0.1099\pm 0.0062$ \cite{wmap}.
The blue (darker) points do not saturate the
relic density $\Omega h^2 \leq 0.0913$ (3$\sigma$  deviation).
The impact of the constraint $m_h\!>\! 114.4$ GeV is also considered.
Compare this figure to Figure~\ref{all} where relic density
constraint was not included. The parameters  values
quoted in eq.(\ref{cmssmparameters}) 
are further restricted, as seen from these plots.
The continuous line is that of minimal 
electroweak $\Delta$ (no relic density constraint).
}\label{allagain}
\end{figure}

It is interesting to see the impact  on $\Delta$ and on the CMSSM parameters
 from  the presence of the dark matter relic density constraint (examined 
using micrOMEGAs2.2 \cite{Belanger:2006is}) and the LEPII constraint on $m_h$.
These are rather strong constraints, particularly in the
``restrictive''  context of CMSSM that we study, with universal
gaugino mass. The results are presented in Figure~\ref{allagain}, where the
relative impact of the LEPII constraint can be seen by comparing the
left and right plots. If the observed  dark matter abundance is imposed as
a constraint on the CMSSM, then the range of values given in
(\ref{cmssmparameters}) and valid for $\Delta<100$
is further restricted, as seen in Figure~\ref{allagain}. 
The condition that the SUSY LSP should provide the observed
dark matter abundance as well as the constraint
$m_h>114.4$ GeV removes the intermediate values of $m_{1/2}$
and  $m_0$, but has a  rather small impact on $A_0$.

\begin{figure}[ht!] 
\begin{tabular}{cc|cr|} 
\parbox{7.3cm}{\psfig{figure=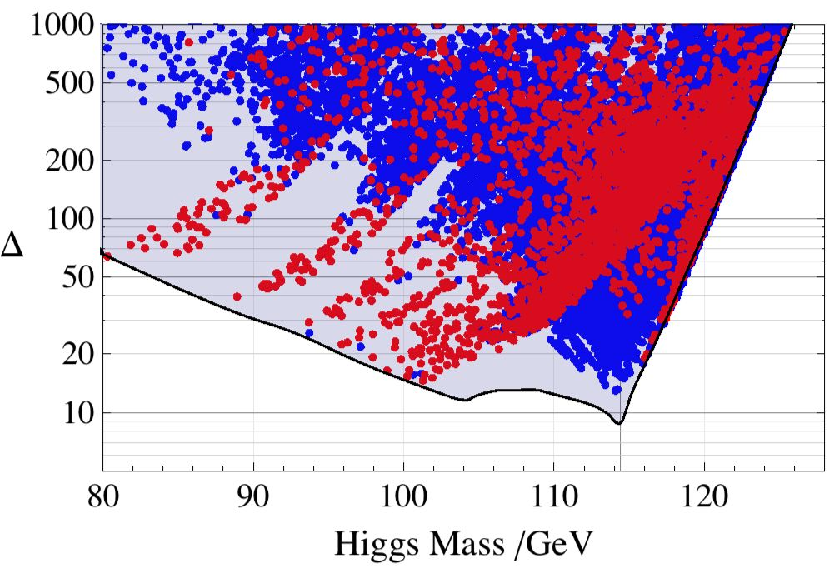,
height=5.5cm,width=6.8cm}} 
\parbox{7.3cm}{\psfig{figure=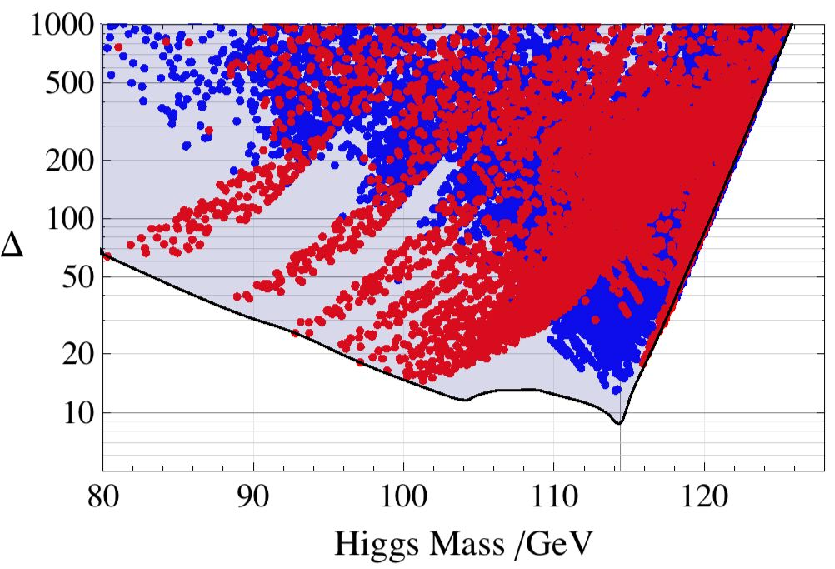,
height=5.5cm,width=6.8cm}}
\end{tabular}
\renewcommand{\baselinestretch}{1.1}
 \caption{\small
Fine tuning vs Higgs mass with the influence of the WMAP bound. The minimum
fine tuning at 2-loop for $2\leq \tan \beta \leq 55$ is given by the
solid line when including all the constraints listed in
Table~\ref{contable}. Left figure: 
The blue (darker) points sub-saturate the
relic density.
 The red (lighter) points give a relic density within the
1$\sigma$ bounds, $\Omega h^2 = 0.1099\pm0.0062$. The `strips' of
points at low Higgs mass appear due to taking steps of 0.5 in $\tan
\beta$ below 10. A denser scan is expected to fill in this
region. Similarly, more relic density saturating points are expected
to cover the wedge of sub-saturating points at
 $m_h^{}\sim114$~GeV and $\Delta \gtrsim 30$. 
Right: as for left, within 3$\sigma$ WMAP bound (in red).
The continuous line is that of minimal electroweak $\Delta$ without the relic density constraint.
}\label{omcon}
\end{figure}

\subsection{Prediction for  $m_h$ from minimising $\Delta$ and
saturating the relic density}\label{mhrelic}

The relic density constraint can be combined with that of minimal
electroweak fine-tuning $\Delta$ to make an interesting prediction for $m_h$.
Figure~\ref{omcon} shows  the impact of non-baryonic relic density
constraint on $\Delta$ presented in Figure~\ref{2loop}. Obviously, not all
initial points in $\Delta$ satisfy this constraint, and
this is shown in Figure~\ref{omcon} by the red and blue points
which do not fill the whole area above the continuous line of
 minimal $\Delta$.  As expected, the additional 
dark matter constraint prefers in some cases
larger $\Delta$ relative to its minimal value (continuous line)
obtained  only with the constraints in Table~\ref{contable}.
However, as can be seen in the plots, the region of $m_h$ 
values where this constraint is indeed relevant 
is actually  ruled out by LEPII bound $m_h>114.4$ GeV; above this
value  the two curves on the boundary are almost 
overlapped and the constraints in
Table~\ref{contable} are sufficient to also satisfy the thermal relic
density; note that the red points in the left (right) plots in 
Figure~\ref{omcon} satisfy the relic density within 1$\sigma$ and 3$\sigma$ 
WMAP bounds \cite{wmap}, respectively. The results are obtained 
as usual, using two-loop values for the quartic couplings 
and soft masses and corresponding threshold effects.
(with the SOFTSUSY and micrOMEGAs codes).

It is important to notice that,
{\it without} imposing the LEPII bound, at the
two-loop level, the smallest fine tuning $\Delta$ consistent with
the relic density WMAP bounds \cite{wmap} 
predicts a mass for the lightest Higgs
as follows:
\bea\label{mhfromrd}
m_h&=&114.7\pm 2 \,\,\,\mbox{GeV}, \,\,\,\,\,\Delta=15.0,\,\,\,\,\,
\textrm{(sub-saturating the WMAP bound).}\nonumber\\
m_h&=& 116.0\pm 2 \,\,\,\mbox{GeV}, \,\,\,\,\,\Delta=19.1,\,\,\,\,\,
\textrm{(saturating the WMAP within 1$\sigma$).}\nonumber\\
m_h&=&115.9\pm 2 \,\,\,\mbox{GeV}, \,\,\,\,\,\Delta=17.8,\,\,\,\,\,
\textrm{(saturating the WMAP within 3$\sigma$).}
\eea

\smallskip
To conclude, minimising the fine-tuning together with the constraints
from precision electroweak data,
 the bounds on SUSY masses and  the requirement of the 
observed dark matter abundance lead to a prediction for $m_h$,
 without imposing the LEPII bound.
This is an interesting result, and represents
our prediction for the CMSSM lightest Higgs mass based on assuming
$\Delta$ as a quantitative test of SUSY as a solution to
the hierarchy problem.

\section{Dark matter fine tuning and its effect on the Higgs mass}

The dark matter abundance can be very sensitive to the choice of
parameters and can introduce a new fine tuning to the model.
To quantify this it is interesting to consider the dark matter fine tuning 
$\Delta^\Omega$ wrt the CMSSM parameters, and to determine
 its impact on the
overall fine tuning (for earlier studies see 
\cite{Chankowski:1998za,Ellis:2007by} and references therein).
 Its definition is similar to that of $\Delta$:
\medskip
\bea
\Delta^\Omega=
\max\bigg\vert\frac{\partial \ln \Omega h^2}{\partial \ln q}
\bigg\vert_{q=m_0,m_{1/2},A_0,\tan\beta}
\eea

\medskip
\begin{figure}[t!] 
\begin{tabular}{cc|cr|} 
\parbox{7.3cm}{\psfig{figure=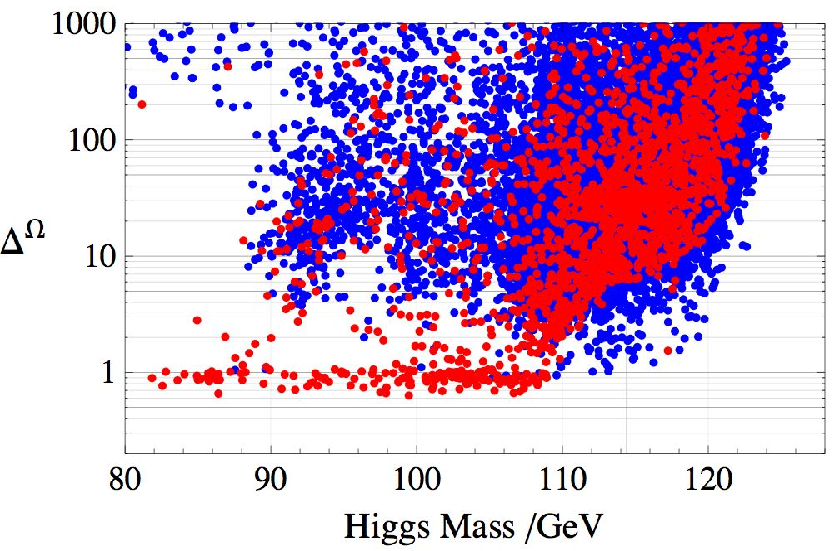,
height=5.5cm,width=6.8cm}} 
\parbox{7.3cm}{\psfig{figure= 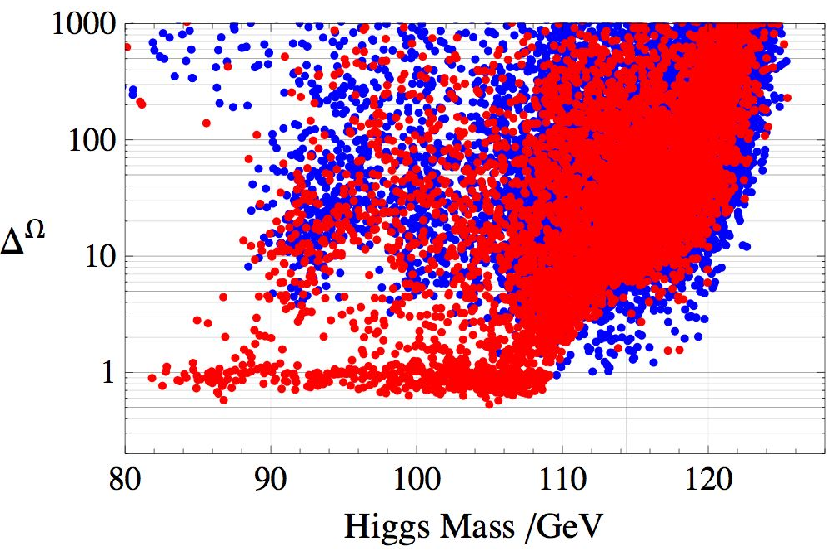,
height=5.5cm,width=6.8cm}}
\end{tabular}
\renewcommand{\baselinestretch}{1.1}
\caption{\small
Left (Right) figure:
Relic density fine tuning, $\Delta^\Omega$ in function of
the Higgs mass, at two-loop level,
for a 1$\sigma$  (3$\sigma$)  WMAP bound (red), respectively.
The blue (darker) points sub-saturate the
dark matter relic density.
}
\label{ft1}
\end{figure}

\begin{figure}[t!] 
\begin{tabular}{cc|cr|} 
\parbox{7.3cm}{\psfig{figure=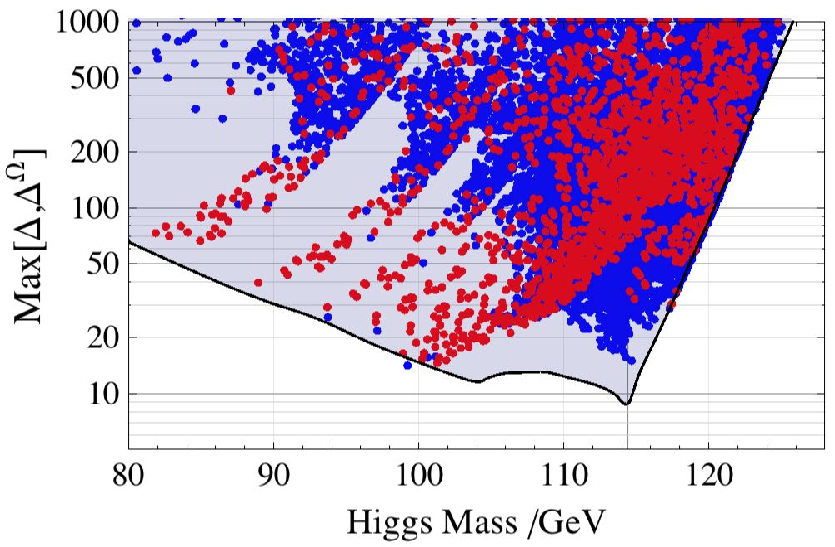,
height=5.5cm,width=6.8cm}}
\parbox{7.3cm}{\psfig{figure= 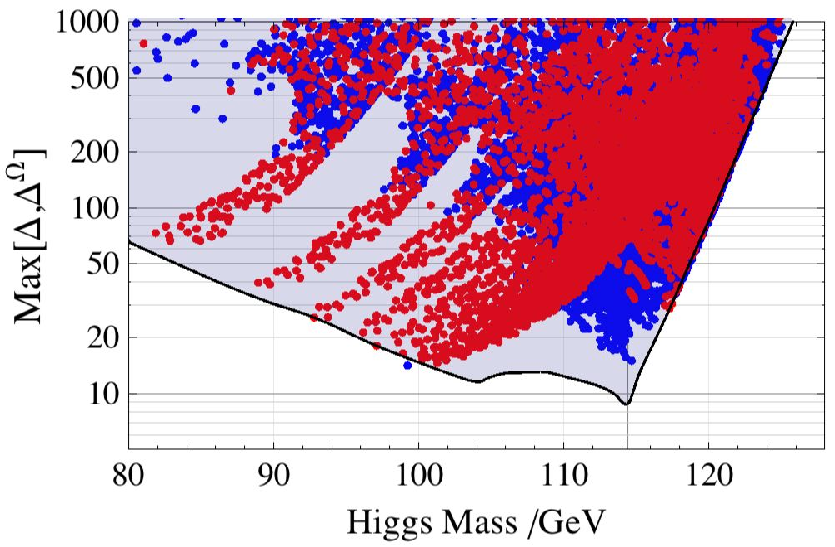,
height=5.5cm,width=6.8cm}}
\end{tabular}
\renewcommand{\baselinestretch}{1.1}
\caption{\small
Left (Right) figure:
The total fine tuning, $\max\{\Delta,\Delta^\Omega\}$ in function of
the Higgs mass, at two-loop level,
for a 1$\sigma$  (3$\sigma$)  WMAP bound (in red), respectively.
The blue (darker) points sub-saturate the
dark matter relic density.
The continuous line represents the minimal value of the EW fine-tuning
computed earlier.
}
\label{ft2}
\end{figure}

\medskip 
In  Figure~\ref{ft1} we evaluated $\Delta^\Omega$ at two-loop level and
presented as a function of the Higgs mass, without imposing any
restriction on the latter. It turns out that $\Delta^\Omega$ can have
acceptable values even for $m_h$ close to $120$ GeV.
In Figure~\ref{ft2} the total fine-tuning, defined as
$\max\{\Delta,\Delta^\Omega\}$ is presented as a function of $m_h$.
Its value is only slightly larger than that found earlier 
for $\Delta$ alone with WMAP saturated dark matter abundance
(in red in the plots). From Figure~\ref{ft2} we predict, from minimising
 $\max\{\Delta,\Delta^\Omega\}$ and from
 consistency with the 3$\sigma$ WMAP bound:
\medskip
\bea\label{mhfromrdft}
m_h&=&\!\!114.70\pm 2\,\,\mbox{GeV},\,\,
\max\{\Delta,\Delta^\Omega\}=15,\,\,\,\,\,\,
\textrm{(sub-saturating WMAP bound)}\nonumber
\\
m_h&=&\!\!116.98\pm 2\,\,\mbox{GeV},\,\,
\max\{\Delta,\Delta^\Omega\}=28.7,\,\,\,
\textrm{(saturating WMAP bound within 3$\sigma$)\,\,\,\,\,\,}
\eea

\medskip\noindent
The last predicted  value of $m_h$ is only marginally above that predicted in
(\ref{mhfromrd}), 
based on minimised electroweak fine-tuning and right dark matter
abundance.

\section{Predictions for the superpartners 
 from fine tuning limits}

The results so far demonstrate that electroweak fine-tuning
has a strong sensitivity to parameters such
as  $\mu$, $m_{1/2}$, with a preference for lower values.
Regarding the $m_0$ dependence, $\Delta$ has a rather flat
dependence when we are in the focus point region. 
The states that are dominantly controlled by the 
$\mu$, $m_{1/2}$  parameters are then the most important in determining
the naturalness of the proposed theory. These include the neutralinos,
charginos and the gluino states. 
Further, setting an upper bound on electroweak $\Delta$
 gives a bound on the spectrum.
If any of these states have
masses in excess of those given in Table~\ref{partlimit}, it will
require less than 1\% tuning ($\Delta>100$) for the MSSM.

\medskip
\begin{table}[ht]
\begin{center}
\begin{tabular}{|c||c|c|c|c||c|c||c|c||c|c|}
\hline
$\tilde{g}$ & $\chi_{1}^{0}$ & $\chi_{2}^{0}$ & $\chi_{3}^{0}$ & $
\chi_{4}^{0}$ & $\chi_{1}^{\pm}$ & $\chi_{2}^{\pm}$ & $\tilde{t}_{1}^{}$ & $
\tilde{t}_{2}^{}$ & $\tilde{b}_{1}^{}$ & $\tilde{b}_{2}^{}$ \\ \hline\hline
1720 & 305 & 550 & 660 & 665 & 550 & 670 & 2080 & 2660 & 2660 & 3140 \\ 
\hline
\end{tabular}
\end{center}
\caption{\small 
Upper mass limits on superpartners in GeV such that $\Delta<100$
remains possible. }
\label{partlimit}
\end{table}

These upper mass limits scale approximately as
$\sqrt{\Delta_{\mbox{\tiny min}}^{}}$, so they may be adapted depending on
how much fine tuning the reader is willing to accept. 
Overall low fine tuning prefers 
a Higgsino mass of $\cO(0.5~\textrm{TeV})$,
a gluino of $\cO(1.5~\textrm{TeV})$ and chargino and neutralino masses of
$\cO(300~\textrm{GeV})$. Stop and sbottom masses are significantly
larger at $\cO(3~\textrm{TeV})$ 
due to the weak limit on $m_0$ (focus point).

Finally we return to the intriguing fact that minimum electroweak
fine tuning plus
correct dark matter abundance corresponds to a Higgs mass just above
the LEPII bound\footnote{One may ask  whether the fine-tuning measure 
used above  has indeed a clear physical meaning. One can object that
nature may not choose  ``minimal'' fine-tuning results.
 One can invoke the example of
{\it classical} chaotic systems, displaying the familiar
``butterfly effect'' where small variations
of the initial conditions bring large changes
of the final state (``fine tuning''), yet the system is ``realised''. 
Such effects exist in (nonlinear) 
classical systems, where initial close values (states)
of a parameter exponentially diverge after evolving according to the
dynamics of the differential eqs. 
In our setup, one could have such
effects not from evolution in time but from evolution wrt the 
energy scale, from the high scale to the low scale, after
including  {\it quantum} effects encoded in the RG differential equations.
By this analogy 
one could object that using criteria of low fine-tuning to obtain
 mass bounds (for $m_h$) may not be appropriate. However,
the difference is that the discussion in the text is at 
{\it quantum level}, so the counterexamples of {\it classical}
(chaotic) systems do not necessarily apply.}.
As we noted above this point is fixed by the current bounds on the SUSY
spectrum and not by the current Higgs mass bound which is not included when
doing the scans giving Figs~\ref{2loop},
\ref{dplot}, \ref{omcon}, \ref{ft2}.

One may interpret the SUSY parameters
corresponding to this point as being the most likely given our present
knowledge and so it is of interest to compute the SUSY spectrum for this
parameter choice as a benchmark for future searches. This is presented in
Table~\ref{favspec2} where it may be seen that it is somewhat
 non-standard with very
heavy squarks and sleptons and lighter 
neutralinos, charginos and gluinos. This has  similarities
to the SPS2 scenario \cite{BA}. 

\begin{table}[th]
\begin{center}
\begin{tabular}{|c|c||c|c||c|c||c|c|}
\hline
$h^{0}$ & 114.5 & $\tilde{\chi}_1^0$ & 79 & $\tilde{b}_{1}^{}$ & 1147 & $
\tilde{u}_{L}^{}$ & 1444 \\[1pt] 
$H^{0}$ & 1264 & $\tilde{\chi}_2^0$ & 142 & $\tilde{b}_{2}^{}$ & 1369 & $
\tilde{u}_{R}^{}$ & 1446 \\[3pt] 
$H^\pm$ & 1267 & $\tilde{\chi}_3^0$ & 255 & $\tilde{\tau}_{1}^{}$ & 1328 & $
\tilde{d}_{L}^{}$ & 1448 \\[2pt] 
$A^0$ & 1264 & $\tilde{\chi}_4^0$ & 280 & $\tilde{\tau}_{2}^{}$ & 1368 & $
\tilde{d}_{R}^{}$ & 1446 \\[2pt] 
$\tilde{g}$ & 549 & $\tilde{\chi}_1^\pm$ & 142 & $\tilde{\mu}_L^{}$ & 1406 & 
$\tilde{s}_{L}^{}$ & 1448 \\[2pt] 
$\tilde{\nu}_{\tau}^{}$ & 1366 & $\tilde{\chi}_2^\pm$ & 280 & $\tilde{\mu}
_R^{}$ & 1406 & $\tilde{s}_{R}^{}$ & 1446 \\[2pt] 
$\tilde{\nu}_{\mu}^{}$ & 1404 & $\tilde{t}_{1}^{}$ & 873 & $\tilde{e}_L^{}$
& 1406 & $\tilde{c}_{L}^{}$ & 1444 \\[2pt] 
$\tilde{\nu}_{e}^{}$ & 1404 & $\tilde{t}_{2}^{}$ & 1158 & $\tilde{e}_R^{}$ & 
1406 & $\tilde{c}_{R}^{}$ & 1446 \\ \hline
\end{tabular}
\end{center}
\par
\caption{\small The favoured
 Constrained MSSM spectrum of minimal $\Delta = 15$
giving a sub-saturation of the WMAP bound. Masses are given in $GeV$.}
\label{favspec2}
\end{table}

\subsection{Predictions for SUSY searches at the LHC}

It is clear that there is still a wide range of parameters that needs to be
explored when testing the CMSSM. Will the LHC be able to cover the whole
range? To answer this note that, for a fine tuning measure
$\Delta<100$, one must be able to exclude the upper limits
of the mass parameters appearing in Table~\ref{partlimit}.
Of course the state
that affects fine tuning most is the Higgs scalar and one may see from
Figure~\ref{2loop}
 that establishing the bound $m_{h}>120$\,GeV will imply that 
$\Delta>100.$ However the least fine tuned region corresponds to the lightest
Higgs consistent with the LEPII bound and this is the region where the LHC
searches rely on the $h\rightarrow\gamma\gamma$ channel which has a small
cross section and will require some $30\,fb^{-1}$ at $\sqrt{s}=14$\,TeV to
explore. Given this it is of interest to consider to what extent the direct
SUSY\ searches will probe the low fine tuned regions. 
Following the discussion in the previous section, the most significant
processes at the LHC will be those looking for gluinos, winos and
neutralinos.

Studies of SUSY\ at the LHC~$\cite{Baer:2009dn}$ have shown that the LHC
experiments have a sensitivity to gluinos of mass $1.9$\thinspace TeV for 
$\sqrt{s}=10$\thinspace TeV, $2.4$\thinspace TeV for $\sqrt{s}=14$\thinspace
TeV and luminosity $10fb^{-1}.$ Of relevance to the first LHC\ run the limit
is $600$\thinspace GeV for $\sqrt{s}=10$\thinspace TeV and luminosity 
$100\,pb^{-1}.$ These correspond to probing up to $\Delta =120,180$ and $14$
respectively. 
As we have discussed charginos and neutralinos can be quite light, but their
signal events are difficult for LHC to extract from the background, owing in
part to a decreasing $M_{\widetilde{W}}-M_{\widetilde{Z}}$ mass gap as 
$\left\vert \mu \right\vert $
decreases~\cite{Barbieri:1991vk,Baer:2004qq}. An Atlas study
\cite{Vandelli:2007zza}  of the trilepton signal from
chargino-neutralino production found that $30fb^{-1}$ luminosity at
14 TeV is needed for a
 3$\sigma$ discovery significance
 for $M_2<300$~GeV and $\mu<250$~GeV \cite{CMS}.

\section{Summary and Conclusions}

Supersymmetry was introduced to solve the hierarchy
problem and to avoid the large  fine-tuning in the
SM Higgs sector associated with the Planck or GUT scale
 when quantum corrections are included.
 While this hierarchy problem is solved
by TeV-scale supersymmetry, the non-observation, so far,
 of SUSY states means
that the MSSM has acquired some residual amount of fine-tuning 
related to  unnatural cancellations in the SUSY breaking sector.
The goal of this paper was to analyse in detail the level of
fine tuning in the CMSSM.

The fine tuning measure $\Delta$  provides a quantitative
test of SUSY as a solution to the ``little''
hierarchy problem and measures the  ``tension'' required to satisfy
the scalar potential minimum condition  $v^2\sim -m_{susy}^2/\lambda$, for  
a combination of soft masses $m_{susy}^2\sim$ TeV, with an effective quartic
coupling $\lambda< 1$ and $v\sim \cO(100)$ GeV.
Although the 
exact upper limit on the fine tuning $\Delta$ beyond which a theory
fails to solve the hierarchy problem is debatable,
it is preferable, for a given model, to have a parameter space
configuration corresponding to the lowest value of $\Delta$. We evaluated $\Delta$ at two-loop order and 
also paid particular attention to threshold
corrections and to the $\tan\beta$ radiative dependence on 
the parameters. Such effects on  fine-tuning were not fully considered in the past and turned out to reduce fine
tuning significantly.

Our determination of the  fine-tuning measure for the CMSSM included the theoretical constraints (radiative EWSB,
avoiding charge and colour breaking vacua), and also the 
experimental constraints (bounds on superpartner masses,
electroweak precision data, $b\ra s\,\gamma$, $b\ra \mu\,\mu$ and
muon anomalous magnetic moment, dark matter abundance). As far as we are aware,
our study is the first two-loop analysis of the fine tuning problem
in the CMSSM, largely  based on SOFTSUSY and micrOMEGAs, 
SuSpect and our own Mathematica code. The latter was very important
since it  reduced to a feasible level the CPU run time necessary to scan the full parameter space.

Not including the dark matter constraint, we found the minimum value is given by $\Delta=8.8$. Remarkably, even without imposing the LEP bound on the Higgs mass, the condition fine tuning should be a minimum predicts
$m_h=114\pm 2$ GeV. Adding the constraint on the dark matter relic density, one finds
$\Delta=15$ corresponding to $m_h=114.7\pm 2$ GeV and this rises to 
$\Delta=17.8$ ($m_h=115.9\pm 2$ GeV) for SUSY dark matter abundance 
within 3$\sigma$ of the
WMAP constraint.  The results are encouraging for the search for SUSY
because we considered the ``conservative'' case of CMSSM, 
and it is well-known that relaxing gaugino universality  
can reduce $\Delta$ further \cite{Kane:1998im, Horton:2009ed}.

The spectrum corresponding to the minimum value of the fine tuning shows similarities to the SPS2  scenario with light
neutralinos, charginos and gluinos (corresponding to light $\mu$,
$m_{1/2}$) and heavy squarks and sleptons corresponding to
large $m_0$, near the focus point limiting value \cite{Feng:2000bp,Chan:1997bi}. It provides the "best" estimate for the SUSY spectrum given the present  experimental bounds.

Increasing $m_h$ above the minimum fine tuned value causes $\Delta$ 
to increases exponentially fast and one leaves 
the focus point region
at the edge of which this minimal value is
reached; one obtains $\Delta~\!=~\!100\,\,(1000)$ for a scalar mass
$m_h=121$ ($126$) GeV, respectively. Ultimately the question whether the SUSY solution to the hierarchy
problem has been experimentally tested relies on what value
of fine tuning represents the limit of acceptability.
Given a value one can determine the range of parameter space 
that is still acceptable. For the case that the fine tuning measure
should satisfy $\Delta<100$, we determined the corresponding 
superpartners masses and CMSSM parameters values, that can be 
relevant for SUSY searches.

\section*{Acknowledgements}

The research was partially supported by the EU RTN grant UNILHC 23792. S.C. is supported by the UK Science and Technology Facilities 
Council PPA/S/S/2006/04503.
D.G. thanks the Theory Group at \'Ecole Polytechnique 
Paris for their kind hospitality and acknowledges 
the financial support from the ERC Advanced Grant 
ERC-2008-AdG 20080228 (``MassTeV''). D.G. thanks 
S. Pokorski for many interesting discussions on this topic.

\newpage
\section{Appendix}\label{appendixA}

\def\theequation{A-\arabic{equation}}
\def\thesubsection{A}
\setcounter{equation}{0}
\def\thefigure{A-\arabic{figure}}

\subsubsection{Higgs mass and EW fine tuning}

We provide technical results used in the  text to evaluate
$\Delta$. The potential used in (\ref{2hdm})
\medskip
\bea
V&=&  m_1^2\,\,\vert H_1\vert^2
+ m_2^2\,\,\vert H_2\vert^2
- (m_3^2\,\,H_1 \cdot H_2+h.c.)\nonumber\\[3pt]
 &&
 ~+~
\frac{1}{2}\,\lambda_1 \,\vert H_1\vert^4
+\frac{1}{2}\,\lambda_2 \,\vert H_2\vert^4
+\lambda_3 \,\vert H_1\vert^2\,\vert H_2\vert^2\,
+\lambda_4\,\vert H_1\cdot H_2 \vert^2\nonumber\\
 && ~+~
\bigg[\,
\frac{1}{2}\,\lambda_5\,\,(H_1\cdot  H_2)^2+\lambda_6\,\,\vert H_1\vert^2\, 
(H_1 \cdot H_2)+
\lambda_7\,\,\vert H_2 \vert^2\,(H_1 \cdot H_2)+h.c.\bigg]
\label{2hdm2}
\end{eqnarray}

\medskip\noindent
with $H_1.H_2=H_1^0\,H_2^0-H_1^- H_2^+$.
Using the notation
\medskip
\begin{eqnarray}
m^2 &=&
 m_1^2 \, \cos^2 \beta +  m_2^2
 \, \sin^2 \beta - m_3^2 \, \sin 2\beta\nonumber\\[3pt]
\lambda &=&\frac{ \lambda_1^{} }{2} \, \cos^4 \beta 
+ \frac{ \lambda_2^{} }{2} \,  \sin^4 \beta 
+ \frac{ \lambda_{345}^{} }{4} \, \sin^2 2\beta 
+ \sin 2\beta \left( \lambda_6^{} \cos^2 \beta 
+ \lambda_7^{} \sin^2 \beta \right)\label{ml2}
\end{eqnarray}

\medskip\noindent
the minimisation conditions give
\medskip
\bea
v^2=-m^2/\lambda,\qquad 
2 \lambda\frac{\partial m^2}{\partial \beta}=m^2\frac{\partial
  \lambda}{\partial \beta}
\eea
or, equivalently,
\medskip
 \begin{eqnarray}
 \frac{2m_3^2}{\sin 2\beta} ~=~ m_1^2 + m_2^2 + \frac{v^2}{2} \left[
   \lambda_1^{} \, c^2_\beta + \lambda_2^{} \, s^2_\beta + 
 \lambda_{345}^{} + \left( \lambda_6^{}+ \lambda_7^{} \right)  s_{2\beta}^{} 
 +\lambda_6^{} \cot \beta + \lambda_7^{} \tan \beta \right] 
 \label{B0fix}
\nonumber \\
 m_1^2 - m_2^2 \tan^2 \beta ~=~ -\frac{v^2}{2} \left[ \cos^2 \beta
   \left( \lambda_1^{} - \lambda_2^{} \tan^4 \beta \right) +
  \sin 2\beta \left( \lambda_6^{} - \lambda_7^{} \tan^2 \beta \right) \right]
 \label{mu0fix}
 \end{eqnarray}
One finds for the CP odd Higgs mass:
\medskip
\begin{eqnarray}
m_A^2 &=& \frac{2 m_3^2}{\sin 2\beta } -\frac{ v^2 }{2} 
\, \left( 2 \lambda_5^{} +  \lambda_{6}^{} \cot \beta + 
\lambda_7^{} \tan \beta  \right),
\hspace{10mm}
m_Z^2 ~=~ \frac{g^2 \, v^2}{4}
\end{eqnarray}

\medskip\noindent
with $g^2=g_1^2+g^2_2$. In the notation of \cite{Davidson:2005cw},
the CP even Higgs masses are
\medskip
\begin{eqnarray}
m_h^2 &=& \frac12 \Big[ \, m_A^2 + v^2 \left( 2 \lambda +
 \Lambda_5^{} \right) - \sqrt{ \left[ m_A^2 + v^2 \left( \Lambda_5^{}-
 2 \lambda \right) \right]^2 + 4 v^4 \, \Lambda_6^2\,} \, \Big]
\nonumber\\
\Lambda_5^{} &=&
\frac{s^2_{2\beta} }{4} \, \left( \lambda_1^{} + 
\lambda_2^{} -2 \lambda_{345}^{} \right) + \lambda_5^{}
 - \frac{s_{4\beta}^{}}{2} \,(\lambda_6^{}-\lambda_7^{})
\nonumber\\[3pt]
\Lambda_6^{} &=&  \frac{s_{2\beta}^{} }{2} \left( \lambda_{3451}^{}
 \,c^2_{\beta}- \lambda_{3452}^{} \,s^2_{\beta} \right) 
+ \frac{c_{2\beta}^{}}{2}   \left( \lambda_6^{} + \lambda_7^{} \right)
+ \frac{c_{4\beta}^{}}{2}   \left( \lambda_6^{} - 
\lambda_7^{} \right)
\end{eqnarray}

\medskip\noindent
where $\lambda_{345j}^{} = \lambda_{345}^{}-\lambda_j^{}$,
$\lambda$ is that of (\ref{ml}), and, $(s_\beta, c_\beta) = (\sin \beta , \cos \beta)$. In the
limit of large $m_A^{}$, $m_h^2$ reaches an upper limit of $2\lambda
v^2$ (which tends to $\lambda_2^{} v^2$ for large $\tan \beta$). At tree level,  $\lambda = (g^2/8) \cos^2 2 \beta$, 
and the individual $\lambda_j^{}$ are:
\begin{eqnarray}
\lambda_{1,2}^{} ~=\,  - \lambda_{345}^{} ~=~ \frac{g^2}{4}
\hspace{15mm}
\lambda_{5,6,7}^{} ~=~ 0
\end{eqnarray}

\medskip\noindent
The general formula for fine-tuning $\Delta_p$ wrt a parameter $p$
in a two-higgs doublet model that we are using can be found in the Appendix
of \cite{Cassel:2009ps}. For large $\tan\beta=v_2/v_1$ this reduces to
\medskip
\bea
\Delta_p^{} &=& \frac{\partial \ln v^2}{\partial \ln p}
\nonumber\\[3pt]
&=& 
\frac{
[ 4 \lambda_7\,(m_3^2)' - 4 \lambda_7' \,m_3^2 ] 
+ [ \lambda_2^\prime \, v^2 +  2 ( m_2^2)' ]  \, [ \lambda_{3452} 
+ 2 ( m_1^2-  m_2^2)/v^2 ]  }{2 \lambda_7^2 \,v^2 - \lambda_2 
\,[ \lambda_{3452}\,\,v^2 + 2\,( m_1^2- m_2^2) ] }
+{\mathcal O}(\cot\beta)
\\[3pt]
&\ra&
- \frac{1}{\lambda_2^{} \, v^2} \left[
 2\, (m_2^2)'  +  \lambda_2^\prime \, v^2  
  +
  4\, v^2
\left(
\frac{
   \lambda_7\,(m_3^2)' -  \lambda_7' \,m_3^2  
 }{  \lambda_{3452}\,\,v^2 + 2\,( m_1^2- m_2^2)  }
\right)
\right],\,\,\,
(\textrm{if}\,\,\lambda_7\vert\ll\vert\lambda_{2}\vert,\vert
\lambda_{3452}\vert)\nonumber
\label{fttan}
\end{eqnarray}

\medskip\noindent
where $x^\prime = \partial x / \partial \ln p$ is the partial
 derivative of $x$ wrt $p$.

\bigskip\bigskip
\subsubsection{The scalar potential: 1 Loop Leading Log (1LLL) Terms}

Here we show how to obtain the one-loop corrected potential, which is
 ``improved'' to the two-loop leading log (2LLL) result in the next section.
Start with
\medskip
\begin{eqnarray}\label{v0}
V^{(0)} &=&  
\bar{m}_1^2 \left| H_1^{} \right|^2 + \bar{m}_2^2 \left| H_2^{}
 \right|^2 - \bar{m}_3^2 \left( H_1^{} H_2^{} + \hc \right)
+ \frac{g^2}{8} \left( \left| H_1^{} \right|^2 - \left| H_2^{}
 \right|^2 \right)^2
\end{eqnarray}

\medskip\noindent
This receives (field dependent threshold) corrections, 
computed using the 
Coleman-Weinberg potential \cite{Coleman:1973jx}:
\begin{eqnarray}
V^{(1)} &=& \frac{1}{64 \pi^2} \, \sum_k^{} \, 
 (-1)^{2 J_k^{}} \, (2 J_k^{}+1)\, g_k^{} 
\, m_k^4 \left( \log \frac{m_k^2}{Q^2} - \frac32 \right)
\label{cwpotential}
\end{eqnarray}
where $m_k^{}$ is the field dependent mass, the degeneracy factor
 $g_k^{}$
 is 6 for squarks, and $J_k^{}$ is the particle spin. All parameters
 in eq.(\ref{cwpotential}) are evaluated at the scale Q using the RGEs
 which ignore the particle 
masses. The field dependent squark masses are 
(neglecting $O\! \left( g^4 \right)$ terms): 
\medskip
\begin{eqnarray}
m_{\tilde{t}_{1,2}^{}}^2 &\approx& M_S^2 + h_t^2  \left| H_2^{} \right|^2  
+ \frac{g^2}{8} \left( \left| H_1^{} \right|^2 - \left| H_2^{} \right|^2  \right)
\mp  h_t^{} 
 \left| A_t^{} H_2^{} - \mu H_1^* \right|
\\
m_{\tilde{b}_{1,2}^{}}^2 &\approx& M_S^2 + h_b^2  \left| H_1^{} \right|^2  
+ \frac{g^2}{8} \left( \left| H_2^{} \right|^2 - \left| H_1^{} \right|^2  \right)
\mp h_b^{} 
\left| A_b^{} H_1^{} - \mu H_2^* \right|
\end{eqnarray}

\medskip\noindent
and where $m_{Q,U,D}^{} (M_S^{}) = M_S^{}$ is assumed. Here, $M_S^{}$
is  the soft SUSY breaking squark mass evaluated at the squark mass scale.

One can expand the non-linear field dependence (log) in $V^{(1)}$ in
inverse powers of $1/M_S$ to
find the dominant threshold corrections, which 
 come from the third generation squarks:
\medskip
\begin{eqnarray}
V^{(1)}_{\tilde{t}_{1,2}^{}} &\approx& \frac{3}{16\pi^2} \Bigg[ 
 t \left( h_t^4 \left|H_2^{}  \right|^4 + 2h_t^2 \, M_S^2 \left|
 H_2^{} 
 \right|^2 + h_t^2 \left| A_t^{} H_2^{} - \mu H_1^* \right|^2 \right)
\nonumber \\
&& \hspace{15mm}
+~ h_t^4 ~ \frac{\left| A_t^{} H_2^{} - \mu H_1^* \right|^2}{M_S^2}
 \left( \left| H_2^{} \right|^2 - \frac{\left| A_t^{} H_2^{} - \mu
 H_1^* 
\right|^2}{12 M_S^2} \right)\nonumber\\
&& \hspace{5mm}
+~ \frac{g^2}{8} \left( \left| H_1^{} \right|^2 - \left| H_2^{} \right|^2 \right)
\left( 2t \, h_t^2 \left| H_2^{} \right|^2 + 2 M_S^2 \left( t -1
 \right) + 
\frac{\left| A_t^{} H_2^{} - \mu H_1^* \right|^2}{M_S^2}  \right)
\Bigg]\qquad
\end{eqnarray}
\begin{eqnarray}
V^{(1)}_{\tilde{b}_{1,2}^{}} &\approx& \frac{3}{16\pi^2} \Bigg[ 
 t \left( h_b^4 \left|H_1^{}  \right|^4 + 2h_b^2 \, M_S^2 \left|
 H_1^{}  
\right|^2 + h_b^2 \left| A_b^{} H_1^{} - \mu H_2^* \right|^2 \right)
\nonumber\\
&& \hspace{15mm}
+~ h_b^4 ~ \frac{\left| A_b^{} H_1^{} - \mu H_2^* \right|^2}{M_S^2} 
\left( \left| H_1^{} \right|^2 - \frac{\left| A_b^{} H_1^{} - \mu
 H_2^* 
\right|^2}{12 M_S^2} \right)
\nonumber\\
&& \hspace{5mm}
+~ \frac{g^2}{8} \left( \left| H_2^{} \right|^2 - \left| H_1^{} \right|^2 \right)
\left( 2t \, h_b^2 \left| H_1^{} \right|^2 + 2 M_S^2 \left( t -1
 \right) + 
\frac{\left| A_b^{} H_1^{} - \mu H_2^* \right|^2}{M_S^2}  \right)
\Bigg]\qquad\label{lam}
\end{eqnarray}
where 
\bea
t = \log (M_S^2/Q^2)
\eea

\medskip\noindent
 When running below the EWSB scale,
the inclusion of higher dimensional terms (threshold corrections) lead to a
re-summation such that $M_S^{}$ is replaced by a mass scale related to
the physical particle masses \cite{Carena:1995wu}.
 For the results of this paper,
the geometric mean of the stop masses is used in the place of $M_S^{}$.

The above equations are valid down to the top mass scale; below this
 scale threshold corrections from the top quark should also be included. The
 dominant effect of running below the top scale can be absorbed by
 setting $Q$ in the above equations as the ``running'' top mass
 evaluated at the scale $Q$ instead of the pole mass.

From eqs.(\ref{v0}) to (\ref{lam}) one obtains the 
parameters entering in the scalar potential (\ref{2hdm}), evaluated at
the scale Q (below $M_S$), in the one-loop leading log approximation (1LLL):
\medskip
\begin{eqnarray}
m_1^2 &=& \bar{m}_1^2 - \frac{6h_b^2}{16 \pi^2} \, M_S^2  
+ \frac{3 }{16 \pi^2}  \left( 2 h_b^2 \, M_S^2 + h_b^2 A_b^2 
+ h_t^2 \mu^2 \right) t  \label{m1loop1}  \\[3pt]
m_2^2 &=&  \bar{m}_2^2 - \frac{6h_t^2}{16 \pi^2} \, M_S^2 
 + \frac{3 }{16 \pi^2}  \left( 2 h_t^2 \, M_S^2 + h_t^2 A_t^2
 + h_b^2 \mu^2 \right) t  \\[3pt]
m_3^2 &=& \bar{m}_3^2 + \frac{3 }{16 \pi^2} \left( h_t^2 A_t^{}
 + h_b^2 A_b^{} \right) \mu\, t 
 \label{lambda1}
\end{eqnarray}
\begin{eqnarray}
\lambda_1^{} &=& \frac{g^2}{4} \left( 1 +  \frac{3 \left( h_t^2\, 
\mu^2 - h_b^2\, A_b^2 \right)}{16 \pi^2 \,  M_S^2} \right) + 
\frac{3}{8 \pi^2} \left( 
 \frac{h_b^4 \, X_b^{}}{2} 
- \frac{h_t^4 \, \mu^4}{12 M_S^4} \right)
+ \frac{3h_b^2 }{8 \pi^2} \left( h_b^2 - \frac{g^2}{4} \right) t 
\hspace{10mm} \\[10pt]
\lambda_2^{} &=& \frac{g^2}{4} \left( 1 + \frac{3 \left( h_b^2\, 
\mu^2 - h_t^2\, A_t^2 \right) }{16 \pi^2 \,  M_S^2} \right) + 
\frac{3}{8 \pi^2} \left( 
 \frac{h_t^4 \, X_t^{}}{2} 
- \frac{h_b^4 \, \mu^4}{12 M_S^4} \right)
+ \frac{3h_t^2 }{8 \pi^2} \left( h_t^2 - \frac{g^2}{4} \right) t  
\eea
\bea
\lambda_{34}^{} &=& -\frac{g^2}{4}  \left( 1 +  \frac{ 3 h_t^2 
\left( \mu^2 - A_t^2 \right)  }{32\pi^2 \,  M_S^2} +  
\frac{ 3 h_b^2 \left( \mu^2 - A_b^2 \right) }{32\pi^2 \,  M_S^2} \right)
+ \frac{3 \left( h_t^2 + h_b^2 \right) }{16 \pi^2}\, 
\frac{g^2}{4}  \, t \nonumber \\[5pt]
&& \hspace{10mm}
+~ \frac{3 h_t^4}{16\pi^2} \left( \frac{ \mu^2 }{M_S^2} - 
\frac{  \mu^2 A_t^2 }{3 M_S^4} \right)
+ \frac{3 h_b^4}{16\pi^2} \left( \frac{ \mu^2 }{M_S^2} - 
\frac{  \mu^2 A_b^2 }{3 M_S^4} \right) \qquad \qquad  \quad
\eea
\bea
\lambda_5^{} &=& 
- \frac{3 h_t^4}{96\pi^2}   \frac{\mu^2 A_t^2}{M_S^4}
- \frac{3 h_b^4}{96\pi^2}  \frac{\mu^2 A_b^2}{M_S^4}  \\[10pt]
\lambda_6^{} &=& 
\frac{g^2}{4} \left( \frac{3 \mu \left( h_b^2 A_b^{} - h_t^2 A_t^{} 
\right)}{ 32 \pi^2 \, M_S^2} \right)
+ \frac{3 h_t^4}{96\pi^2} \, \frac{\mu^3 A_t^{}}{ M_S^4} 
+ \frac{3 h_b^4}{96\pi^2} \, \frac{\mu}{M_S^{}} 
\left( \frac{ A_b^3}{ M_S^3} - \frac{ 6 A_b^{}}{M_S^{}}  \right) 
\eea
\bea
\lambda_7^{} &=& 
\frac{g^2}{4} \left( \frac{3 \mu \left( h_t^2 A_t^{} - h_b^2 A_b^{} 
\right)}{ 32 \pi^2 \, M_S^2} \right)
+ \frac{3 h_b^4}{96\pi^2} \, \frac{\mu^3 A_b^{}}{ M_S^4} 
+ \frac{3 h_t^4}{96\pi^2} \, \frac{\mu}{M_S^{}} 
\left( \frac{ A_t^3}{ M_S^3} - \frac{ 6 A_t^{}}{M_S^{}}  \right)
\label{lambda7}
\end{eqnarray}

\medskip\noindent
These analytic results agree with \cite{Carena:1995bx} which ignore
 the stop mixing corrections to the D-terms, but are included
 here for completeness. The following notation is used in this appendix.
\medskip
\begin{eqnarray}
X_{t,b}^{} &=& \frac{2 A_{t,b}^2}{M_S^2} \left( 1- 
\frac{A_{t,b}^2}{12 M_S^2} \right)
\end{eqnarray}

\subsubsection{The scalar potential: 2 Loop Leading Log (2LLL) Terms}

The two-loop leading log (2LLL) Coleman-Weinberg potential 
can be found in the arXiv version of \cite{Espinosa:2000df}
to $O\! \left( g_3^2 \, h_t^4 , g_3^2 \, h_b^4 \right)$ and $O\!
\left( h_t^6, h_t^4 \, h_b^2, h_t^2 \, h_b^4, h_b^6 \right)$,  see
also \cite{Martin:2001vx,Martin:2002iu} for the general case.
 The method of
the previous section may be used to determine the 2LLL contributions
to the Higgs scalar potential, however here we proceed instead with an
approach similar to that in
 \cite{Carena:1995wu}, to RG-improve the 1-loop result into a 2LLL result.
A step approximation is applied to the $\beta$-functions so that the
MSSM RG eqs are used between the GUT and stop mass scale, 
then the 2HDM SM RG eqs between the stop and top mass scales, and finally
the top is integrated out to reach the electroweak scale.

When setting the renormalisation scale in eqs~(\ref{m1loop1}) to
 (\ref{lambda7}) as $Q=M_S^{}$, the logarithmic terms are removed but
 the finite corrections from stop mixing remain. These results are
 then used as boundary conditions for the parameters at the scale
 $M_S^{}$. A series  expansion of the RG eqs is then applied:
\medskip
\begin{eqnarray}
\lambda \left( Q \right) &\approx& \lambda \left( M_S^{} \right) 
- \beta_\lambda^{} \left( M_S^{} \right) \, t + \frac12 \, 
\beta^\prime_\lambda \left( M_S^{} \right) \, t^2 + O\! \left( t^3 \right) \\
&=& \lambda \left( M_S^{} \right) - \beta_\lambda^{} 
\left( Q \right) \, t - \frac12 \, \beta^\prime_\lambda 
\left( Q \right) \, t^2 + O\! \left( t^3 \right)
\label{rgimp}
\end{eqnarray}
where $\beta_p^{} = \partial p / \partial \log Q^2$.
Eventually, all parameters will be expressed at a scale $Q$ as 
in the Coleman-Weinberg potential approach. For a
 $\beta_{\lambda}^{}$-function of the form
 $b \, \lambda + c$, eq~(\ref{rgimp}) becomes
\medskip
\begin{eqnarray}
\lambda  &\approx& \lambda \left( M_S^{} \right) - 
t \left[ b \, \lambda \left( M_S^{} \right) + c  \right]
+ t^2 \left[ b \,c  - \frac12 \, \beta^{\prime}_{\lambda} + O\! 
\left( \lambda \right) \right]
\label{lambdarg}
\end{eqnarray}

\medskip\noindent
where the couplings are evaluated at the scale $Q$ unless stated
 otherwise. The $\beta$-functions for the 2HDM SM \cite{Haber:1993an}
 are  listed below, 
neglecting $O\! \left( h_\tau^2 \right)$ terms, and
 with the $\beta_{\lambda_i^{}}^{}$-functions also neglecting
 $O\! \left( g^4 ,  g^2  \lambda _i^{} , \lambda_i^2 \right)$ terms:
\medskip
\begin{eqnarray}
16 \pi^2 \, \beta_{m_{1}^2}^{} &=&
3 h_b^2 \, m_1^2  + O\left( g^2 m^2 \right)\nonumber  \\
16 \pi^2 \, \beta_{m_{2}^2}^{} &=&
3 h_t^2 m_2^2 + O\left( g^2 m^2 \right)\nonumber \\
16 \pi^2 \, \beta_{m_{3}^2}^{} &=&
\frac32 \left( h_t^2 +h_b^2 \right) m_3^2  + O\left( g^2 m^2 \right)
\end{eqnarray}
\begin{eqnarray}
16 \pi^2 \, \beta_{\lambda_1^{}}^{} &\approx&
6 h_b^2 \left( \lambda_1^{}  - h_b^2 \right)\nonumber  \\
16 \pi^2 \, \beta_{\lambda_2^{}}^{} &\approx&
6 h_t^2 \left( \lambda_2^{} - h_t^2 \right) \nonumber \\
16 \pi^2 \, \beta_{\lambda_3^{}}^{} &\approx&
3\lambda_3^{}  \left( h_t^2 + h_b^2 \right) 
-6 h_t^2 h_b^2 \nonumber \\
16 \pi^2 \, \beta_{\lambda_4^{}}^{} &\approx&
3\lambda_4^{} \left( h_t^2 + h_b^2 \right)  
+6 h_t^2 h_b^2  \nonumber \\
16 \pi^2 \, \beta_{\lambda_5^{}}^{} &\approx&
3 \lambda_5^{} \left( h_t^2 + h_b^2 \right) \nonumber \\
16 \pi^2 \, \beta_{\lambda_6^{}}^{} &\approx&
 \lambda_6^{} \left( \frac92 \, h_b^2 + \frac32 \, h_t^2 \right)
\nonumber \\
16 \pi^2 \, \beta_{\lambda_7^{}}^{} &\approx&
 \lambda_7^{} \left( \frac92 \, h_t^2 + \frac32 \, h_b^2 \right) 
\end{eqnarray}
and finally
\begin{eqnarray}
16 \pi^2 \, \beta_{h_t^2}^{} &\approx&
h_t^2 \left(
\frac92 \, h_t^2 + \frac12 \, h_b^2 - 8 g_3^2 
- \frac94 \, g_2^2 -\frac{17}{12} \, g_1^2 \right) \\
16 \pi^2 \, \beta_{h_b^2}^{} &\approx&
h_b^2 \left(
\frac92 \, h_b^2 + \frac12 \, h_t^2 + h_\tau^2
- 8 g_3^2 - \frac94 \, g_2^2 -\frac{5}{12} \, g_1^2 \right) 
\end{eqnarray}

\medskip
Using (\ref{lambdarg}), the analytic 2-loop results in
\cite{Carena:1995bx} are then recovered when the same level
 of approximation is considered. For example,
\medskip
\begin{eqnarray}
\lambda_2^{} &\approx& 
\left[ \lambda_2^{} \left( M_S^{} \right) 
-\lambda_2^{} \, a_2^{} \, t \right]  - b_2^{} \, t
+  \left[ \, a_2^{} \, b_2^{}  + \frac{3 h_t^2}{16\pi^2 }
 \left( 2 \beta_{h_t^2}^{} - \beta_{\lambda_2^{}}^{}  \right)
 + O\! \left(  \lambda \right)  \right]  t^2
\\[2pt]
&=& \left[ \lambda_2^{} \left( M_S^{} \right)
 -\lambda_2^{} \, \frac{6h_t^2}{16\pi^2}  \, t \right] 
+ \frac{3 h_t^4}{8\pi^2} \left[ t + \frac{1}{16 \pi^2}    
\left( \frac32 \, h_t^2 + \frac12 \, h_b^2 - 8 g_3^2 \right)  t^2 \right]
\end{eqnarray}

\medskip\noindent
The couplings entering  in the expression of $\lambda_2^{} \left(
M_S^{} \right)$ are re-expressed in terms of their values
 at the scale $Q$, (with a
 logarithmic correction which compensates for the running below $M_S^{}$):
\medskip
\begin{eqnarray}
h_t^4 (M_S^{} ) &=& h_t^4 \left( 1 + \frac{t}{16\pi^2} \left( 9  h_t^2
+  h_b^2 - 16 g_3^2 \right) + O\! \left( g^2 \,t, t^2 \right) \right)
\label{YukRunT} 
\\[3pt]
h_b^4 (M_S^{} ) &=& h_b^4 \left( 1 + \frac{t}{16\pi^2} 
\left( 9  h_b^2 + 
 h_t^2 - 16 g_3^2 \right) + O\! \left( g^2 \, t, t^2 \right) \right)
\label{YukRunB}
\end{eqnarray}

\medskip\noindent
This
leads to the following expression,
 in agreement with \cite{Carena:1995bx},
 when the stop mixing contributions to the D-terms in the 
potential are neglected:
\medskip
\begin{eqnarray}
\lambda_2^{} &\approx&
\frac{g^2}{4} \left( 1- \frac{3 h_t^2}{8 \pi^2} \, t \right) 
-  \frac{3 h_b^4}{96 \pi^2} \, \frac{\mu^4}{M_S^4} \left[ 1 + 
\frac{t}{16\pi^2} \left( 9h_b^2-5 h_t^2 - 16 g_3^2 \right) \right]
\nonumber \\[10pt]
&& +~ \frac{3 h_t^4}{8\pi^2} \left[ t +\frac{X_t^{}}{2} + 
\frac{t}{16 \pi^2} \left( \frac{3 h_t^2}{2} + \frac{h_b^2}{2} 
- 8 g_3^2 \right) \left( X_t^{} + t \right) \right] 
\end{eqnarray}

\medskip\noindent
Note that these results assume that the CP odd Higgs mass is not
decoupled. If this is the case, the usual SM $\beta$-functions
should be used. The effective quartic coupling at the EW scale when $m_A^{} 
\lesssim M_S^{}$ is given by:
\medskip
\begin{eqnarray}
\lambda &\approx& \frac{g^2}{8} \cos^2 2\beta 
\Big[ 1 - \frac{3}{16 \pi^2} 
 \left( h_b^2+h_t^2+ \left( h_b^2 - h_t^2 \right) 
\sec 2\beta \right) t \Big]
\nonumber \\[4pt]
&&
+~ \frac{3 \, h_t^4 }{16\pi^2}\, \sin^4 \beta \bigg[ t+
 \frac{\tilde{X}_t^{}}{2} \, +\frac{1}{16 \pi^2} \left( \frac{3\,
 h_t^2}{2} + \frac{h_b^2}{2} - 8 g_3^2
 \right) \left( \tilde{X}_t^{} \, t+t^2  \right) + \delta_1^{} \bigg]
\nonumber \\[4pt]
&&
+~ \frac{3 \, h_b^4 }{16\pi^2}\, \cos^4 \beta \bigg[ t+
 \frac{\tilde{X}_b^{}}{2} +\frac{1}{16 \pi^2} \left( \frac{3\,
 h_b^2}{2} + \frac{h_t^2}{2} - 8 g_3^2
 \right) \left( \tilde{X}_b^{} \, t+t^2  \right) + \delta_2^{} \bigg]
\end{eqnarray}

\medskip\noindent
with the following notation:
\medskip
\begin{eqnarray}
\delta_1^{} &=&  \frac{3t \left( h_b^2 - h_t^2 \right)}{16 \pi^2}  ~
 \frac{\tilde{A}_t^{} \, \mu \cot \beta}{M_S^2} \, \left( 1-
 \frac{\tilde{A}_t^2 }{6 M_S^2}
 \right)  \\
\delta_2^{} &=&  \frac{3t \left( h_t^2 - h_b^2 \right) }{16 \pi^2} ~
\frac{\tilde{A}_b^{} \, \mu \tan \beta}{M_S^2} \, \left( 1-
\frac{\tilde{A}_b^2 }{6 M_S^2}
 \right)
\end{eqnarray}

\medskip\noindent
where $\tilde{X}_{t,b}^{}$ is defined as $X_{t,b}^{} (A_{t,b}^{} \to 
\tilde{A}_{t,b}^{} )$ with
\medskip
\begin{eqnarray}
\tilde{A}_t^{} &=& A_t^{} - \mu \cot \beta \nonumber\\
\tilde{A}_b^{} &=& A_b^{} - \mu \tan \beta
\end{eqnarray}

\medskip\noindent
A similar but distinct result is obtained when $m_A^{} \sim M_S^{}$ 
(notably no $\delta_i^{}$ terms and a different dependence on $\tan
\beta$ and the mixed Yukawa couplings). The threshold corrections are
dependent on where the CP odd Higgs decouples. The same procedure has
been applied to determine the 2LLL threshold corrections to the mass
terms.

\end{document}